\documentclass{aa}  

\usepackage{graphicx}    
\usepackage{txfonts}    
\usepackage[breaklinks=true,colorlinks=true,linkcolor=OliveGreen,citecolor=Blue,filecolor=red,urlcolor=CadetBlue]{hyperref} %
\usepackage{amsmath}    
\usepackage{amssymb}     
\usepackage{twoopt}
\usepackage{placeins}    
\usepackage{multirow}
\usepackage{siunitx}
\usepackage{float}
\usepackage{booktabs}
\usepackage{mathrsfs}
\usepackage{euscript}
\usepackage[dvipsnames]{xcolor}
\usepackage{orcidlink}

\bibpunct{(}{)}{;}{a}{}{,} 

\newcommand{\MBH}{M_{{\rm BH}}}
\newcommand{\Msun}{M_\odot}
\newcommand{\Min}{M_{\rm in}}

\newcommand{\Rin}{R_{\rm in}}
\newcommand{\Rsat}{r_{\rm sat}}
\newcommand{\Ucr}{u_{\rm cr}}

\makeatletter
\newcommand{\@adsurl}[1]{\href{#1}{ADS}}
\makeatother

\providecommand{\apjl}{ApJL}
\providecommand{\apjs}{ApJS}
\providecommand{\aap}{A\&A}

\nolinenumbers
\let\linenumbers\relax

\begin{document} 

   \title{Iron K$\alpha$ signatures from accretion disks around \\ fermionic dark matter cores}

   \author{V. Crespi
          \inst{1,2} \orcidlink{0009-0005-8190-3598}
          \thanks{valentinacrespi@fcaglp.unlp.edu.ar}
          \and
          F.L. Vieyro \inst{2,3} \orcidlink{0000-0001-9173-0209}
          \thanks{fvieyro@fcaglp.unlp.edu.ar}
          \and
           C.R. Argüelles \inst{1,2,4,5}  \orcidlink{0000-0002-5862-8840} 
          \thanks{carguelles@fcaglp.unlp.edu.ar} 
          \and
          J.A. Rueda \inst{4,5,6} \orcidlink{0000-0003-4904-0014}
          \thanks{jorge.rueda@icra.it}
          }

   \institute{Instituto de Astrofísica de La Plata, UNLP $\&$ CONICET, Paseo del Bosque, B1900FWA La Plata, Argentina
         \and
         Fac. de Ciencias Astron\'omicas y Geof\'isicas, UNLP, Paseo del Bosque, B1900FWA La Plata, Argentina.
         \and Instituto Argentino de Radioastronom\'ia, CONICET-CICPBA-UNLP, CC5 (1894) Villa Elisa, Prov. de Buenos Aires, Argentina
         \and ICRANet, Piazza della Repubblica 10, 65122 Pescara, Italy
         \and ICRA, Dipartimento di Fisica, Sapienza Università di Roma, Piazzale Aldo Moro 5, I-00185 Roma, Italy
         \and ICRANet-Ferrara, Dipartimento di Fisica e Scienze della Terra, UNIFE, Via Saragat 1, 44122 Ferrara, Italy 
              }

   \date{Received \today; accepted \today}

 
  \abstract
   {The fluorescent iron line and its broadening due to relativistic effects are excellent probes to study the inner part of an accretion disk and, consequently, the space-time geometry near the compact object.}
   {We investigate the iron K$\alpha$ line profile within the extended Ruffini-Argüelles-Rueda (RAR) model, which describes a fermionic dark matter distribution on galaxy scales from first physical principles. The most general solutions are characterized by a compact and highly degenerate inner core able to mimmic the central black hole, transitioning into an extended and dilute halo composed of the same particles. 
   We aim to contrast the resulting line morphologies in this scenario with those predicted by the standard Kerr black hole paradigm, highlighting their similarities and differences. In particular we choose the iconic Seyfert 1 galaxy MCG-06-30-15 where a broad iron line has been measured, to make a proper comparison.}
   {We compute the iron line profile using the numerical relativistic ray-tracing code, \texttt{Skylight}. We consider two distinct configurations for the emissivity of the cold accretion disk: an irradiation profile based on the lamp-post corona prescription, and a phenomenological power-law profile.}
   {The resulting line profiles exhibit a diverse phenomenology. In particular, the most compact fermion cores produce a line broadening comparable to that observed in rapidly rotating (prograde) black holes. Moreover, because the RAR solution is non-singular, stable circular orbits can extend to very small radii. Consequently, the presence of emitting matter at radii smaller than a gravitational radius $r<r_g$ yields distinctive spectral features that are entirely absent in the black hole scenario. For the specific case of MCG-06-30-15, we find a good agreement with the observed broad features of the iron line profile, provided the compactness of the (non-rotating) fermion core is close to critical.}
   {These results reinforce the need for independent black hole spin measurements. Combined with such constraints, iron-line spectroscopy may provide a powerful observational tool to distinguish black holes from alternative compact solutions, in particular compact fermionic dark-matter cores.}

   \keywords{Black hole physics -- 
            Accretion, accretion disks --
            Line: profiles --
            dark matter --
            Methods: numerical
            }

   \maketitle
%
\section{Introduction}

The nature of supermassive compact objects residing at the centers of galaxies remains a topic of intense investigation. The supermassive black hole (SMBH) paradigm is currently the most widely accepted framework and has been remarkably successful in explaining a broad range of observations, from stellar dynamics around galaxy centers \citep{2024A&ARv..32....3G}, to high-energy emission processes in active galactic nuclei (AGN) \citep{2012ARA&A..50..455F}. However, several open issues remain unsolved, such as the rapid growth of SMBHs seeds in the early Universe, the detailed physics of accretion and feedback and, more fundamentally, the exact nature of these objects. These challenges motivate the exploration of alternative models either within or beyond the framework of General Relativity (GR) .

In this context, the Ruffini-Argüelles-Rueda (RAR) model \citep{Ruffini2015} or its more realistic extension including for particle evaporation \citep{Arguelles2018,Arguelles2019b,Arguelles2021}\,\footnote{This model is also named in the literature as the relativistic fermionic King model \citep{2022PhRvD.106d3538C}.}, offers a novel framework to describe dark matter (DM) halos through self-gravitating, neutral fermions in thermodynamic equilibrium within GR. 
The extended RAR model naturally predicts a continuous density profile consisting of a dense, degenerate core at galactic centers surrounded by a more diffuse halo. 

For fermionic particle masses in the $\mathcal{O}(10-100)$ keV, this unified DM distribution has successfully described various galactic phenomena, such as galaxy rotation curves and galaxy Universal relations \citep{Krut2023}, stellar streams \citep{Mestre2024,Collazo2025}, and the recently measured Keplerian decline of the Milky Way rotation curve \citep{2026PhRvD.113b3010K,Crespi2026}. Simultaneously, the fermionic solutions contain a dense central core capable of mimicking the gravitational influence of a central SMBH for a given particle mass range, as demonstrated by the orbits of the S-stars around Sagittarius A* \citep{2020A&A...641A..34B,Becerra-Vergara2021,2022MNRAS.511L..35A,Crespi2026}. Indeed, by assuming such a dense fermion-core of $\approx 4\times10^6 M_\odot$ lurks at the center of our Galaxy instead of a black hole, we were able to put strict constraints on the fermion mass ($m_f$) in the range of $56-378$ keV \citep{2026PhRvD.113b3010K,Crespi2026}. The upper bound corresponding to the critical core-mass of collapse into a SMBH as the one inferred for SgrA* (see also Sec. \ref{sec:rar}).

There is a broad consensus that the process of accretion onto compact objects (either black holes or not) gives rise to a diverse phenomenology in the emitted luminosity and broadening of the fluorescent iron line \citep{2021SSRv..217...65B}. In a previous work, the standard accretion disk solution \citep{Shakura1973} was explored within the extended RAR spacetime \citep{Millauro2024}, showing that accretion efficiencies can be comparable to those of rapidly rotating Kerr black holes, and that the resulting luminosity spectra can be remarkably similar to those of SMBHs \citep{Millauro2024}. Furthermore, relativistic ray-tracing simulations have shown that the images produced by accretion disks around these fermionic cores present features resembling those observed around SMBHs, such as a central brightness depression and a surrounding ring-like structure, while lacking photon rings, a key distinguishing characteristic \citep{Pelle2024}.

In the standard active galactic nucleus (AGN) paradigm, the hot corona is a key component in shaping the observed wide-band spectrum. Located around the compact central object, its (direct) emission is produced via Comptonization of seed photons from the disk by hot electrons, which results in a characteristic power-law spectrum across the X-ray band. 
A fraction of these energetic photons irradiates the inner regions of the disk, where they are absorbed by iron atoms and subsequently re-emitted through fluorescence. 
This process results in the emission of iron K-shell lines (in the range $6.40-6.97$ keV, depending on the ionization state), with the K$\alpha$ line at $6.4$ keV being the strongest and most commonly observed line in X-ray spectra \citep{Fabian2000}.
 
The theoretical modeling of the iron K$\alpha$ line profile has evolved significantly since the seminal work of \citet{Fabian1989}, who first computed the fully relativistic emission from thin Keplerian disks around Schwarzschild black holes. 
This framework was later extended to the Kerr geometry by \citet{Laor1991} for maximally rotating black holes, and subsequently generalized for arbitrary spin parameters by \citet{Fanton1997}. 
While research has predominantly focused on the standard Kerr paradigm, given the lack of clear evidence abut such a geometry, recent studies have increasingly explored the line's phenomenology in non-standard spacetimes. 
These include investigations into parametric deviations from the Kerr metric \citep{Bambi2013a,Gao2026}, as well as regular solutions and exotic compact objects \citep{Bambi2013b,Bambi2013c,Rosa2024}, providing a versatile tool to test gravity in the strong-field regime. 

In this work, we present general results for the broad iron line profile generated by accretion disks that extends around compact cores described by the extended RAR model. Our goal is to expand the phenomenology of accretion within this framework and to identify potential observational deviations from the standard black hole scenario. In particular, we apply our model to the well-studied Seyfert galaxy MCG–6-30-15, whose broad iron line has been extensively interpreted within the SMBH paradigm \citep[e.g.,][]{Tanaka1995,Fabian2003,Young2005,Brenneman2006,Brenneman2025}. This comparison will allow us to assess the viability of the RAR model as an alternative explanation for the observed spectral features in MCG–6-30-15 and, more broadly, in low mass AGNs where the central object mass can be interpreted as fermion-core alternative to the black hole. Since the estimated mass of the SMBH candidate in MCG–6-30-15 is between (3.6 - 6)$\times 10^6 M_\odot$ \citep{McHardy2005}, similar to our own Galaxy, we will use in this work particle masses within the above stated range for $m_f$ where the fermion core is stable.  

The article is organized as follows: In Sec. \ref{sec:rar} we review the main properties of the RAR solution and compute density profiles for different compactness parameters of the central object. 
In Sec. \ref{sec:rar-disk} we extend on the formalism to compute  thin accretion disks embedded in a general RAR spacetime.
In Sec. \ref{sec:ironLine}, we describe the method used to compute the iron line profile. The results are presented in Sec. \ref{sec:results}. In Sec. \ref{sec:mcg}, we apply our model to MCG–06-30-15. Finally, in Sec. \ref{sec:conclusions}, we summarize our findings and present our conclusions.

\section{Models}

\subsection{Extended RAR model}\label{sec:rar}

In it's original version derived by Ruffini, Argüelles and Rueda in \citeyear{Ruffini2015}, the model consists of self-gravitating fermions (with spin $s=1/2$) in hydrostatic and thermodynamic equilibrium, presenting an alternative approach than the one offered by N-body simulations to describe dark matter halos.
In \citep{Arguelles2018,Arguelles2019a} a more realistic version of this model was developed accounting for particle evaporation, in which the fermionic dark matter particles obey the Fermi–Dirac like distribution function (DF) given by
\begin{equation}\label{eq:DF}
    f_c(\epsilon \leq \epsilon_c) = \frac{2}{h^3}\frac{1-e^{(\epsilon-\epsilon_c)/kT}}{e^{(\epsilon-\mu)/kT}+1}, ~~~~~~~~ f_c(\epsilon > \epsilon_c) = 0
\end{equation}
where an energy cut-off $\epsilon_c$ is taken into account, $\mu$ is the chemical potential, $T$ is the effective temperature, $k$ is the Boltzmann constant and $h$ is the Planck constant. 

The Fermi-Dirac-like DF as given in Eq. (\ref{eq:DF}) can be obtained from a theory of dark matter halo formation based on a maximum entropy production principle \citep{1998MNRAS.300..981C,2026PhRvD.113b3010K} appropriate within the non-linear stages of structure formation. Indeed, by studying the problem of relaxation of collisionless self-gravitating systems on a coarse-grained level, it can be shown that Eq. (\ref{eq:DF}) is a stationary solution of a generalized Fokker-Planck equation for fermions which includes the physics of violent relaxation and particle evaporation \citep{1998MNRAS.300..981C,2026PhRvD.113b3010K}.
This fermionic model presents a new perspective on dark matter halos which does not rely on a phenomenological formula for the DM profiles as in N-body simulations, but it is based on first physical principles with specific insights on the nature and mass of the DM particle candidate. The extended configurations will depend on the fermion mass $m_f$, the dimensionless temperature parameter $\beta_0 = kT_0/(m_fc^2)$ at the origin, the central degeneracy $\theta_0 = \mu_0/kT_0$ and central energy cut-off $W_0=\epsilon_{c,0}/(kT_0)$ parameters. 

Once these are specified, the corresponding 4-parametric equations of state (EoS) at given radius $r$, i.e. the energy density $\rho_r(m_f, \beta_0, \theta_0 , W_0)$ and pressure $P_r(m_f, \beta_0, \theta_0 , W_0)$, are directly obtained as the corresponding integrals of the DF in momentum space. 
We assume these EoS components to be the diagonal part of a stress–energy tensor under a perfect fluid approximation for the self-gravitating fermions \citep{Ruffini1969}, and solve numerically the Einstein field equations within a static and spherically symmetric background. 

The system equations are solved together with the Tolman and Klein thermodynamic equilibrium conditions within general relativity \citep{Tolman1930,1949RvMP...21..531K}, and particle's energy conservation along a geodesic, which ensures finite distributions. 

A more detailed description, as well as the model equations, can be found in \citet{Arguelles2018,Arguelles2021,Crespi2025}.

\begin{figure}[t]
    \centering
    \includegraphics[width=0.9\columnwidth]{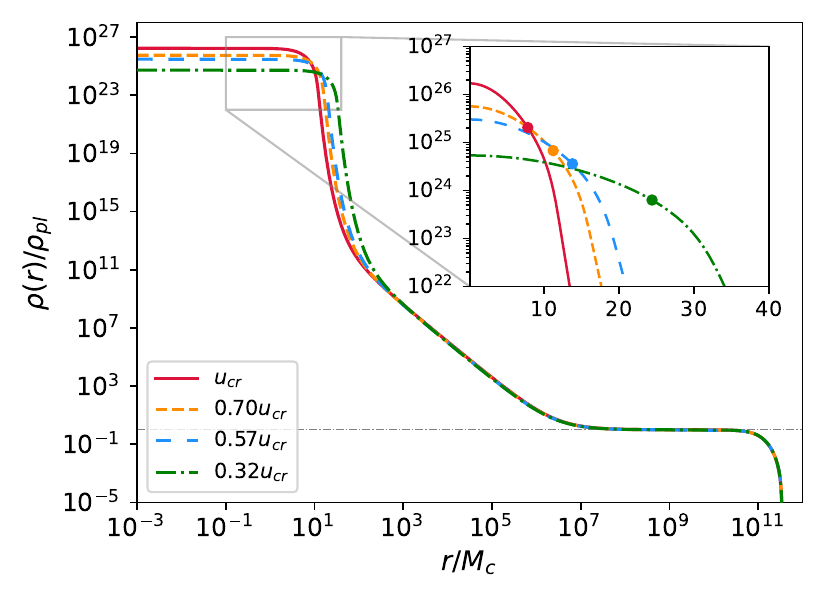}
    \caption{Comparison of the mass density profiles used in this work, in terms of fractional compactness to the critic one $\Ucr=0.128$ (maximum allowed). The zoomed panel shows the locations of the corresponding cores with colored dots. $\rho(r)$ and $r$ are scaled with fixed plateau density $\rho_{pl}$ (gray horizontal line) and core mass $M_c$ values, these are the same for all solutions. The global scale of these core-halo configurations correspond to a Milky Way-like galaxy.}
    \label{fig:rho}
\end{figure}

When considering the quantum effects of the fermions (i.e. through the Pauli exclusion principle) already present in Eq. (\ref{eq:DF}), the resulting morphology consists in a very dense and compact core that is in a highly degenerate state (i.e. $\theta_0 \gtrsim 10$ \cite{Arguelles2021}), which continuously transitions towards the classical (Boltzmannian-like) regime leading to an extended and diluted halo, see Fig. (\ref{fig:rho}). See also \citep{Krut2023} for a detailed comparison with N-body simulations profiles, when contrasted with a large variety of rotation curve observables.

Interestingly, for large enough halos (similar to the Milky Way), these compact cores can achieve masses of the order $\sim 10^6 M_\odot$ (or above, depending on $m_f$), and mimic gravitational signatures of supermassive black holes at their centers \citep{2026PhRvD.113b3010K}\,\footnote{In the case of dwarf spheroidal galaxies, the compact cores can reach masses of the order of $10^4$-$10^6 M_\odot$ and still be surrounded by a fermionic halo able to explain the dispersion velocity \citep{Arguelles2019a}.}. The maximum (critical) core mass $M_c^{cr}$ above which the highly degenerate core collapses towards a black hole do depend on the particle mass, and is given by $M_c^{cr}\approx 0.384\, m^3_{\rm pl}/m_f^2$ \citep{1999EPJC...11..173B,CHAVANIS2020135155,Arguelles2021}, with $m_{\rm pl}$ the Planck mass.  Which, for the particle masses explored in this paper between $300$ and $378$ keV correspond to $M_c^{cr}\approx (4 - 7)\times 10^6 M_\odot$, which is roughly the mass range of the central object inferred for MCG-6-30-15.

Indeed, in \citet{Arguelles2021} a thermodynamic stability analyses of these core-halo configurations was performed, where an exploration of the Turning Point (TP) criterion \citep{Harrison1965,Sorkin1981,2014CQGra..31c5024S} showed that the onset for gravitational instability of the fermionic configurations at finite temperature $T$ is achieved at the maximum of core mass $M_c$ vs. $\rho_0$ central density (i.e. $dM_c/d\rho_0=0$ and $d^2M_c/d\rho_0^2 < 0$). The results in \citep{Arguelles2021} extended the traditional result of Oppenheimer and Volkoff done in the $T=0$ limit \citep{Oppenheimer1939}, to the case of finite temperature configurations within a fully relativistic treatment, with explicit applications to realistic DM halos in a cosmological framework. Moreover, in \citep{Arguelles2021} (and more recently in \cite{2026PhRvD.113b3010K} for Milky Way like galaxies) it was demonstrated that the fermionic core-halo equilibrium solutions close the critical point of SMBH-collapse are thermodynamically stable and extremely long-lived\footnote{Only when sufficient baryonic accretion from high density environments occurs, the dense DM core is driven to the collpse into a SMBH \citep{2024ApJ...961L..10A}.}.

The corresponding compactness parameter $u=M/R$ for these critic configuration --just before gravitational collapse-- takes the value $\Ucr=0.128$, reflecting the core similarity with the Oppenheimer-Volkoff limit for a gas of fully degenerated and relativistic fermions, with a critic compactness of $\Ucr^{OV}=0.125$ \citep{Oppenheimer1939}, both much less than the idealistic upper value given by Buchdahl's theorem of $u<0.44$ \citep{Buchdal1959}.  

In Fig. \ref{fig:rho} we show a comparison of the core-halo family of solutions for the same boundary conditions (same diluted halo and same core mass), but changing the compactness of the core. 
Less compact cores (i.e. lower fractions of $u_{\rm cr}$) translates directly into decreasing the mass of the fermionic particle. Zoomed in, we marked in corresponding dots the location of the core, all sharing the same mass $M_c$. 
This location is defined by the first maximum of the circular velocity curve \citep[see Fig. 1 in][]{Arguelles2019b}.
The mass density profiles are scaled with the plateau density $\rho_{pl}$ defined by the minimum of the circular velocity curve. 
Due to imposing the same boundary conditions $[r_i,M(r_i)]$ for solving the system equations, the value of $\rho_{pl}$ is the same for all solutions. 
The specific values of these compactness are shown in table \ref{tab:table_1}\footnote{Note that changing the central core mass $M_c$ would change the compactness for the same particle mass. Therefore, the critic compactness of $\Ucr=0.128$ is valid for every particle mass, provided the $M_c$ is the critical mass of collapse $M_c^{cr}$.} 
In this work, we focus on the relativistic reflection lines produced in a spacetime background given by the exteded RAR-DM model, characterized by its compactness parameter $u$. We choose all DM cores within this theory to have the same fixed mass $M_c$, while varying the compactness as we change the mass of the fermions. 

\begin{table}[t]
    \centering
    \caption{Compactness considered in this work. All core masses correspond to $M_c = 3.5\times 10^6\ \Msun$. $u/\Ucr$ is the fraction of critic compactness, $\epsilon$ corresponds to radiative energy efficiency, $r_c$, $\Rsat$ are the core radius and saturation radius of the binding energy respectively and $m_fc^2$ is the fermionic particle energy.}
    \begin{tabular}{|l||c c c c|}
    \hline
    $u = M/R$ & $0.128$ & $0.089$ & $0.073$ & $0.041$ \\ \hline
    $u/\Ucr$ & $1.00$ & $0.70$ & $0.57$ & $0.32$ \\
    $\epsilon$ & $0.285$ & $0.191$ & $0.153$ & $0.084$ \\
    $r_c\ [M_c]$ & $7.84$ & $11.23$ & $13.80$ & $24.42$ \\
    $\Rsat \ [M_c]$ & $0.71$ & $0.98$ & $1.20$ & $2.09$ \\ 
    $m_fc^2$ [keV] & $378$ & $365$ & $350$ & $300$ \\
    \hline
    \end{tabular}
    \label{tab:table_1}
\end{table}

\subsection{\texorpdfstring{$\alpha$}{alpha}--disks in the extended RAR potential} 
\label{sec:rar-disk}

We here follow the prescription derived in \citet{Millauro2024} to model steady state cold accretion disks around regular distributions of matter, that extends the seminal accretion disk model by \citet{Shakura1973}. 
By the conservation of mass and angular momentum we have \citep{Millauro2024}:

\begin{align}
    \frac{dm}{dt} &= (-v_r)2\pi r\Sigma, \\
    \eta \Sigma(r) &= \frac{1}{3\pi}\frac{dm}{dt}\left[ 1- \left(\frac{\Min \Rin}{M(r)r}\right)^{1/2}\right] \left[1-\frac{r}{3M(r)}\frac{dM(r)}{dr} \right]^{-1},
\end{align}
where $\Sigma$ is the disk's mass surface density, $v_r$ is a small radial drift velocity, $dm/dt$ is the accretion rate (constant and independent of $r$), $\eta$ is the cinematic viscosity, $M(r)$ is the mass of the background DM configuration, and $\Rin$, $\Min=M(\Rin)$ is the initial radius of the accretion disk and the enclosed mass at that radius. The gas particles of the disk follow nearly Keplerian orbits with an angular velocity given by 
\begin{equation}
    \Omega(r) = \sqrt{\frac{G_N M(r)}{r^3}},
\end{equation}
where $G_N$ is the gravitational constant. 
Due to viscous shear, both faces of the disk will radiate energy away resulting in a viscous dissipation per unit area of
\begin{equation}
    \begin{split}
        D(r) &= \frac{1}{2}\eta\Sigma(r)\left(r\frac{d\Omega(r)}{dr}\right)^{2} \\
        & = \frac{3\Omega^2}{8\pi}\frac{dm}{dt}\left[ 1- \left(\frac{\Min \Rin}{M(r)r}\right)^{1/2}\right] \left[1-\frac{r}{3M(r)}\frac{dM(r)}{dr} \right].
    \end{split}
\end{equation}
In these geometrically thin optically thick accretion disks, it is assumed that every element of the disk radiates as a blackbody, with a local temperature that relates to the viscous dissipation through the Stefan-Boltzmann law \citep{Boltzmann1884} $T(r)= (D(r)/\sigma)^{1/4}$. 
The specific luminosity over the disk results in 
\begin{equation}
    L_\nu = \frac{8\pi h\nu^3}{c^2} \int_{\Rin}^{R_{\rm out}} \frac{r\ dr}{\exp\left[\frac{h\nu}{kT(r)}\right]-1}.
\end{equation}
In \citet{Millauro2024} it was shown that there exists a given core compactness for which the luminosity spectrum is almost indistinguishable from that of a Schwarzschild BH of the same mass $\MBH=M_c$. 

It is well known that extended distributions of matter that are regular solutions of the Einstein field equations, do not posses an innermost (marginally) stable circular orbit (ISCO) \citep{Weinberg1972}. Due to this fact, it is not clear which is the inner radius of an accretion disk embedded in these backgrounds. In the case of solid compact objects (e.g. a neutron star), it is customary to take the surface of the object as the innermost radius of the disk \citep{Shapiro1983,Frank2002}. In transparent (collisionless) matter as the fermionic DM, baryonic particle can penetrate through the core and eventually reach $r \rightarrow 0$. 
In this context, we estimate the inner radius by evaluating the maximum in the gravitational binding energy of massive test particles in circular orbits. For a general metric satisfying stationarity and spherical symmetry, the radiated energy efficiency is given by (taking $c=1$)

\begin{equation} \label{eq:binding-energy}
    \epsilon = 1- \sqrt{g_{00}(r)\left[1+ \frac{r\ g'_{00}(r)}{2g_{00}(r)-rg'_{00}(r)}\right]}.
\end{equation}

\noindent where $g_{00}(r)$ is the time component of the metric tensor and $g'_{00}(r)$ its derivative with respect to the coordinate radius $r$.
In our fermionic solutions, these efficiencies exhibit monotonically decreasing behavior, remaining approximately constant toward the center.  
Consequently, we define a saturation radius $\Rsat$, as the point at which the (inner) efficiency $\epsilon$ varies by less than $1\%$. 
Adopting this definition, the saturation radius typically corresponds to one-tenth of the core radius $\Rsat\ \approx\ 0.1 r_c$; see \citet{Millauro2024} for a more detailed derivation. 

Notably, these fermionic configurations can achieve significantly higher energy efficiencies --without accounting for spin-- than a Schwarzschild BH. For the most compact solution $\Ucr$, the efficiency reaches $28.5\%$, resembling the values of an (astrophysical) maximally rotating Kerr BH \citep{Shapiro1983}. The specific efficiency values and maximum inner disk radius ($\Rsat$) for the compactness used in this work are reported in table \ref{tab:table_1}.

\subsection{Iron K\texorpdfstring{$\alpha$}{alpha} emission line} \label{sec:ironLine}

The main characteristics of the broad iron line shape are well understood by strong and special relativistic effects of the compact central region. 
The intrinsically emitted narrow line --with rest frame energy $E_{K\alpha}=6.4$ keV-- gets modified in the observer's rest frame: i) becomes wider and shifts to redder energies due to gravitational redshift, and ii) becomes broaden into a double-horn shape and skewed due to the relativistic Doppler effect of the moving disk matter, alongside a characteristic asymmetrical profile caused by relativistic beaming that enhances the bluer peak \citep{Fabian2000,Dauser2013}.  
However, there is still great uncertainty in the geometrical configuration of the corona, and consequently, in its emissivity profile impacting the cold accretion disk \citep{Wilkins2012}. 
A wide range of models have been proposed to describe the geometry of the hot coronal gas. 
Some examples include a spherical distribution surrounding the central compact object \citep{Gonzalez2017}, a toroidal distribution in the same plane as the disk between this and the central object \citep{George1991}, and an extended region above and below the disk plane in a \textit{sandwich} configuration \citep{Haardt1991}. 

One of the widely adopted configurations (due to its simplicity) is the lamp-post geometry \citep{Martocchia1996}, where the point-like corona remains stationary at a height $h$ above the disk plane and isotropically irradiates it following a photon emissivity profile $\varepsilon(r)$. 
This can be a practical interpretation for the base of relativistic jets emerging from AGNs \citep{Martocchia1996,Henri1997,Dauser2013}. 
In a flat spacetime, it's easy to see that this profile is proportional to $\varepsilon(r) \propto h/{(r^2+h^2)^{3/2}}$ \citep{Bambi2024}.
When relativistic effects of the central compact object are considered, the gravitational light-bending of photons produce a modified irradiation profile. In the rest-frame of the emitting corona, it's photon spectrum can be well approximated between cut-off energies $E_{\rm min}<E_c< E_{\rm max}$ \citep{Sunyaev1980}, by 
\begin{equation}
    \frac{dN_c}{dE_c dt_c} = K\ E_c^{-\Gamma},
\end{equation}
where $N_c$ is the photon number, $\Gamma$ is the photon index and subscript $_c$ indicates quantities in the corona rest-frame. If one has knowledge on the total luminosity of the corona and it's spectrum, the normalization constant $K$ can be fixed \citep{Bambi2024}.
Since the number of photons is conserved along the rays-path ($dN_c=dN$), the photon specific flux in the disk-frame will be given by
\begin{equation}
    \frac{dN}{dE dt} = K\ {\rm g}^\Gamma\ E^{-\Gamma},
\end{equation}
where photons are redshifted as ${\rm g}=dE/dE_c=dt_c/dt$. 
Therefore, the energy flux that will impinge along the disk results in 
\begin{equation}
    \varepsilon(r) = E\ n(r)\frac{dN}{dE dt} = K\ {\rm g}^\Gamma\ E^{-\Gamma+1}\ n(r),
\end{equation}
where $n(r)$ is the number density of photons. 
This factor carries the geometrical contribution of the emitting corona, and is computed as $n(r)=N(r)/(A(r,dr)\gamma_r)$, i.e., the number of photons $N(r)$ impacting on the disk at radius $r$, over the proper area of an annulus of thickness $dr$ multiplied by the Lorentz factor $\gamma_v$ of a particle in circular orbit in the disk, in order to account for the correct density in the accretion disk frame. 
This incident energy is then reprocessed by the cold accretion disk, and is re-emitted with a profile $I_{\rm e}$ that is (radially) consistent with the emissivity profile of the corona $I_{{\rm e},r} \propto \varepsilon(r)$ in the disk's frame.

In \citet{Dauser2013} a complete relativistic calculation was performed to compute the correct irradiation profile of a hot lamp-post corona upon the accretion disk. There, a detailed analysis in terms of the emissivity index ($\gamma$ in eq. \ref{eq:intensity-PL}) was conducted showing it's radial variation. 
This contrasts with commonly used \textit{ad-hoc} intensity profiles with a power-law behavior \citep{Fabian2002,Bambi2013a,Bambi2013b}
\begin{equation}
\label{eq:intensity-PL}
    I_{{\rm e},r} \propto
    \delta(E-E_{K\alpha})\ r^{-\gamma} .
\end{equation}
%
The authors in \citet{Dauser2013} argue that the emissivity index measured in observations when a power-law irradiation profile is adopted, accounts for the average steepness in a given range of $r$. Therefore, these emissivities can not be properly explained solely by the lamp-post geometry, and other geometric configurations would have to be considered. 

Another contributing factor to the morphology of the fluorescent emission line is the ionization state of the surface layers of the disk. 
These conditions can be quantified with the ionization parameter $\xi(r)$, which is a ratio of  the photo-ionization rate  to the recombination rate, and is directly proportional to the incident X-ray flux $F_X(r)$ along the disk.
According to \citet{Matt1996,Fabian2000}, when $\xi \lesssim 100$ erg cm s$^{-1}$ the disk material is weakly ionized and the reflection spectrum produces an iron emission line at $6.4$ keV. 
This is often refer to as the cold regime, where metal atoms are typically neutral, whereas hydrogen and helium are predominantly ionized. 
However, when $\xi \lesssim 5000$ erg cm s$^{-1}$, a \textit{hot} line occurs at $6.8$ keV, followed by a large absorption edge. When $\xi \gtrsim 5000$ erg cm s$^{-1}$ the disk is highly ionized and there is no iron line production. 
This suggest a certain caution has to be considered when computing the reflection spectrum. 

To compute the flux of the relativistic line observed by a static observer at a distance $D$ of the central compact object and inclination angle $i$ between the normal of the plane detector and the normal axes of the disk plane, we integrate the specific intensity in the observer's rest frame $I_{\rm o}$ over the solid angle $d\Omega_{\rm o}$ subtended by the source
\begin{equation}
    F_{\rm o}(E_{\rm o}) = \int I_{\rm o}(E_{\rm o})\ {\rm cos}(i)\ d\Omega_{\rm o}.
\end{equation}
Since there is no absorption or emission of photons along the ray paths from source to observer, we can relate the intensity measured by the observer $I_{\rm o}$ with energy $E_{\rm o}$ to the intensity measured by the emitter $I_{\rm e}$ with energy $E_{\rm e}$, through the Lorentz invariant quantity \citep{Lindquist1966}
\begin{equation}
    I_{\rm o}(E_{\rm o})\ E_{\rm o}^{-3} = I_{\rm e}(E_{\rm e})\ E_{\rm e}^{-3}.
\end{equation}
Finally, we compute the specific energy flux by integrating over the extent of the disk as
\begin{equation}
    F_{\rm o}(E_{\rm o}) = \frac{2\pi{\rm cos}(i)}{D^2} \int I_{{\rm e},r}(E_{\rm e})\ { g}^3rdr,
\end{equation}
where the factor
\begin{equation}
    g = \frac{E_{\rm o}}{E_{\rm e}} = \frac{p_\mu(B) u^\mu_{\rm o}(B)}{p_\mu(A) u^\mu_{\rm e}(A)},
\end{equation}
encodes the full information of relativistic effects through the four-vector contractions $p_\mu u^\mu$, for a photon emitted at event $A$ with four-momentum $p_\mu$ by the emitter with four-velocity $u^\mu_{\rm e}$, and received at event $B$ by an observer with four-velocity $u^\mu_{\rm o}$ \citep{Weinberg1972}. 

In the next section, we explore the variety of emission lines produced by adopting either a lamp-post corona + $\alpha$-disk configuration, or a phenomenological power-law emissivity profile of the $\alpha$-disk, in a spacetime background produced by a fermionic DM core. 
We also show the contrasts with the well studied Schwarzschild and Kerr BH backgrounds in the literature when changing geometrical and physical parameters of the model.

\section{General results: variety and phenomenology of RAR iron K\texorpdfstring{$\alpha$}{alpha} lines} \label{sec:results}

To numerically compute the broaden emission lines, we use the ray tracing tool \texttt{Skylight} \footnote{\href{https://github.com/joaquinpelle/Skylight.jl}{GitHub Repository: Skylight.jl}} \citep{Pelle2022} fully coded in \texttt{Julia} programming language \citep{Bezanson2017}. This is a robust general-relativistic ray tracing and radiative transfer code, that can be used in arbitrary spacetimes. 

The general setup for the observer rest frame and flux measurement is the same for all computed lines. It consists of it's distance from the central object $D=1000M$, where $M$ refers to the BH mass $\MBH$ or the fermionic DM core $M_c$; in the following we will use this convention when not specified. 
We use the \textit{Pinhole Camera} prescription in \texttt{Skylight}, which allows for arbitrary positions and velocities of the observation point. We choose to measure fluxes along the radial direction in the static frame. We set a resolution of $600$x$600$ pixels for all runs, with a field of view of $3^\circ$x$3^\circ$, capturing the entire source. \\
For the point-like corona setup we assume isotropic emission with a total of $10^6$ photons emerging from $r=h$ along the polar axis, with photon index of $\Gamma=2.0$. 
To compute the emissivity profile that impacts the disk, we take the $50$ radii, $r_i$, between the inner and outer radii of the accretion disk, and bin the null geodesics into small annuli delimited by $r_i$ and $r_{i+1}$. 
The total extent of the disk is $\Delta r=100M$. 
We consider different inclination angles $i$ of the observer's line of sight relative to the rotation axes of the disk. 

\subsection{Comparison with Kerr solution}

\begin{figure*}
    \centering
    \includegraphics[width=0.9\linewidth]{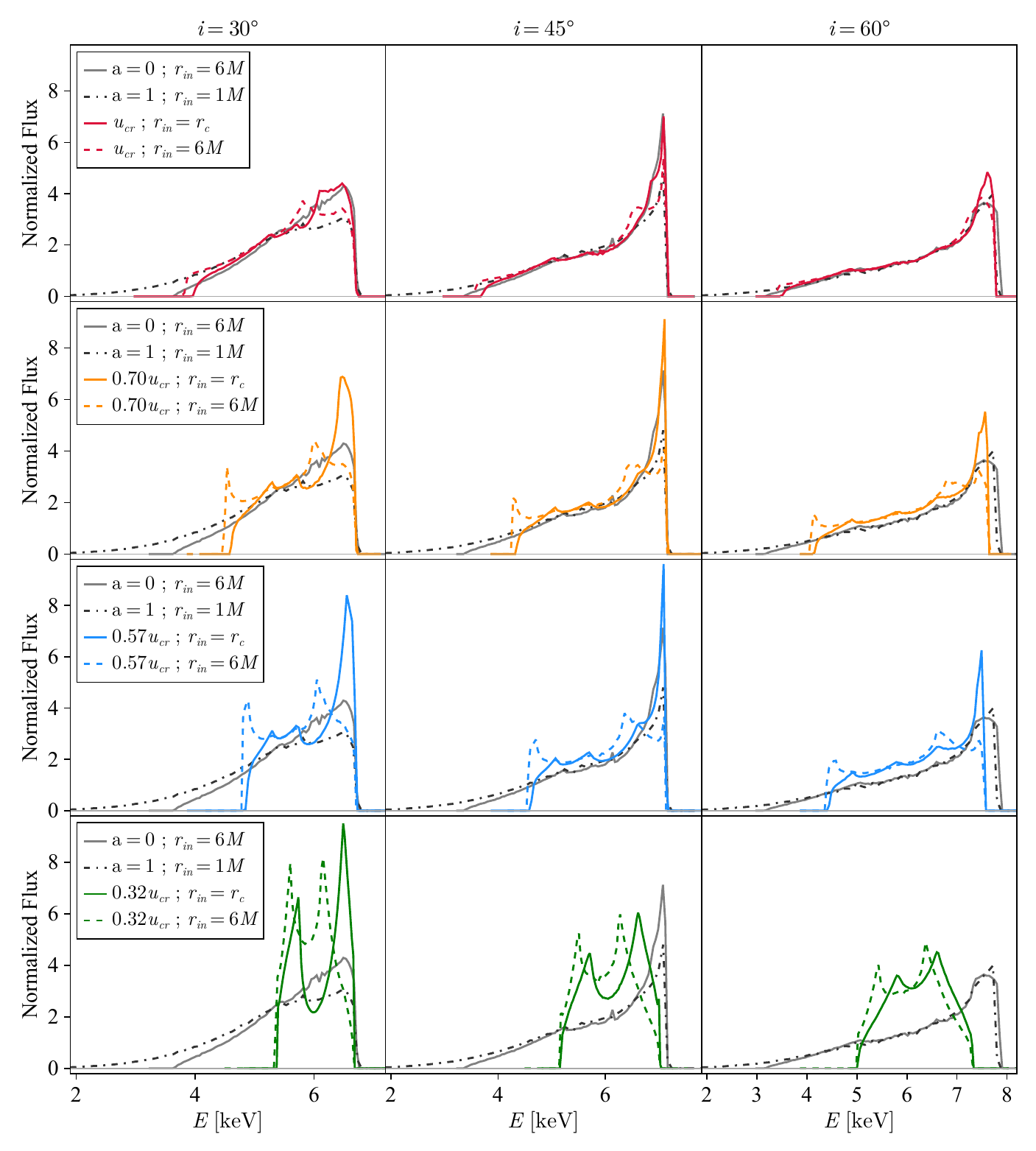}
    \caption{Fe K$\alpha$ emission lines ($E_0 = 6.4$ keV) for fermionic DM models with different compactness. Solid lines correspond to accretion disks with inner radius equal to the core radius $\Rin = r_c$, and dash lines correspond to disks with inner radius equal to $\Rin = 6M_c$. The corresponding values are listed in table (\ref{tab:table_1}).
    From top row to bottom row we decrease the compactness from the critic one. Every column corresponds to different inclinations of the observer's line of sight to the disk axes, from left to right $i=30^\circ, 45^\circ, 60^\circ$.
    As a reference, we show in solid gray the corresponding emission line for a BH with spin $a=0$, and in black dashed line for a BH with spin $a=1$.
    All lines were computed for a lamp-post corona model with $h=5M$ and $\Gamma =2$.}
    \label{fig:mosaic_1}
\end{figure*}

\begin{figure*}
    \centering
    \includegraphics[width=0.95\linewidth]{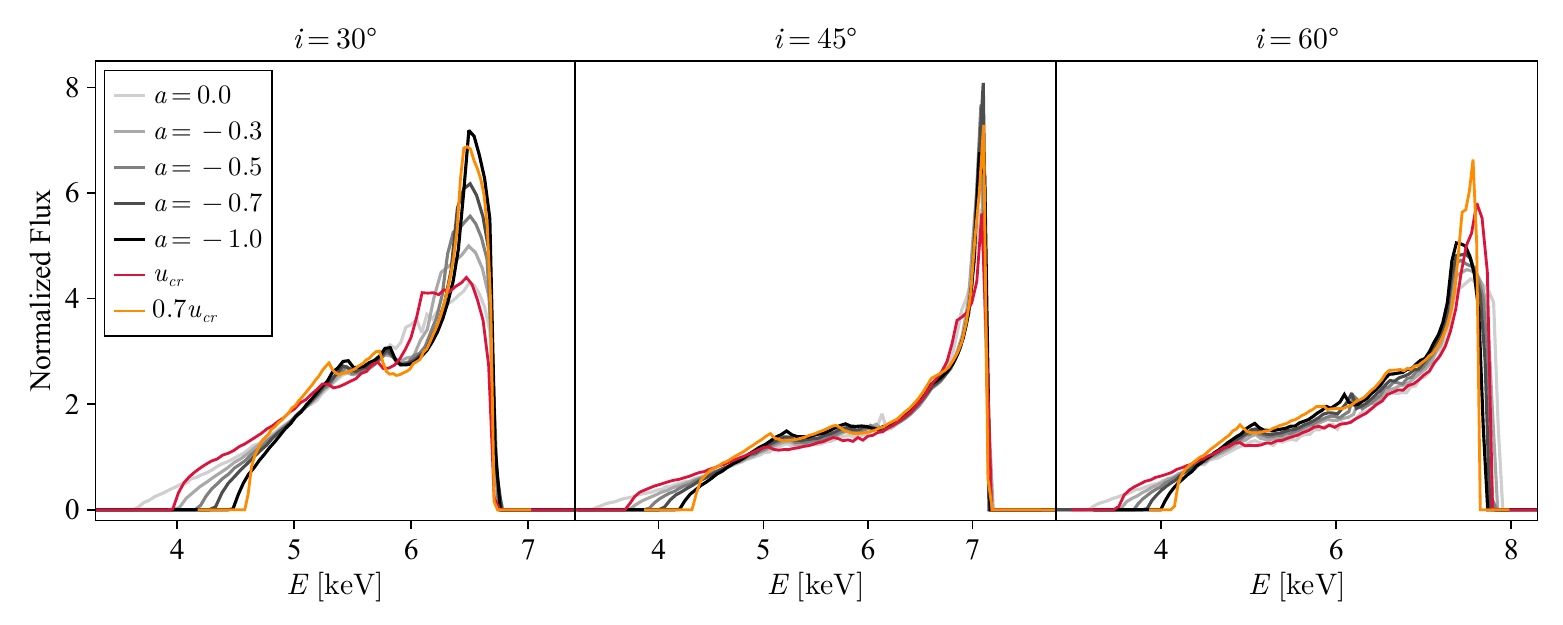}
    \caption{Fe K$\alpha$ emission lines ($E_0 = 6.4$ keV) for Kerr BH with different spin parameter in gray palette, and fermionic DM models with different compactness in colored lines. All inner disk radius correspond to $r_{\rm ISCO}$ of Kerr BH with retrograde rotation sense, and core radius in the case of disks around fermionic cores $\Rin = r_c$. From left to right we increase the inclination angle of the observer's line of sight to the disk axes. All lines were computed for a LP corona model with $h=5M$ and $\Gamma =2$. }
    \label{fig:retrograde_spin}
\end{figure*}

\begin{figure}
    \centering
    \includegraphics[width=0.9\columnwidth]{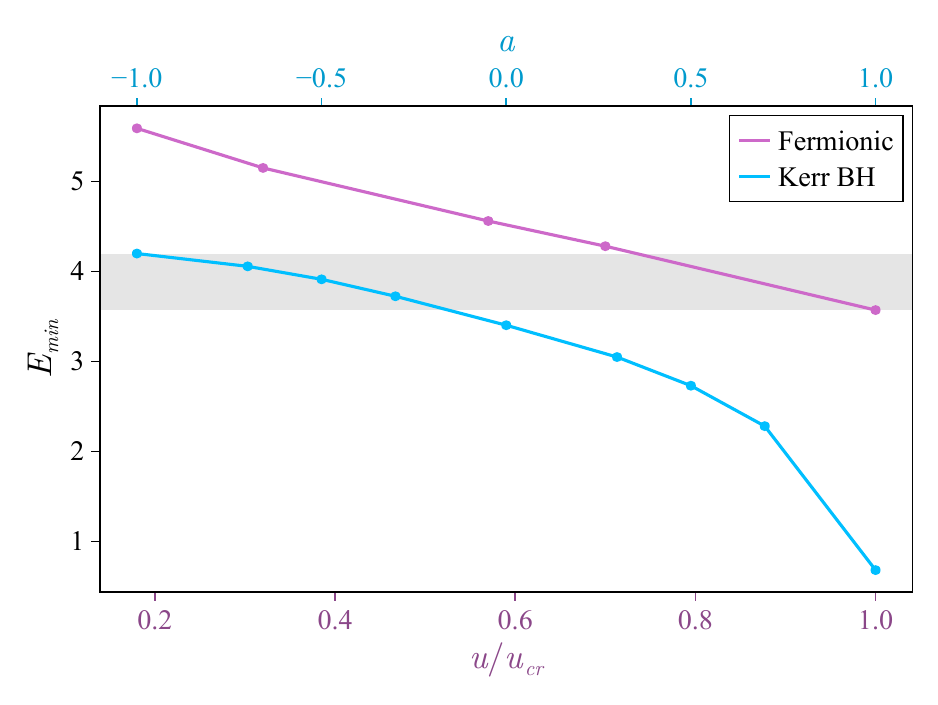}
    \caption{Trends in the red-wing extent, characterized by the onset energy ($E_{min}$ [keV]) of the line profile, at a fixed inclination angle $i=45^\circ$. The blue line corresponds to the top $x$-axis as a function of the spin parameter for Kerr solutions, while the purple line corresponds to the bottom $x$-axis as a function of the compactness fraction for the fermionic RAR-DM model.}
    \label{fig:E_min_comparison}
\end{figure}

\begin{figure}
    \centering
    \includegraphics[width=0.9\columnwidth]{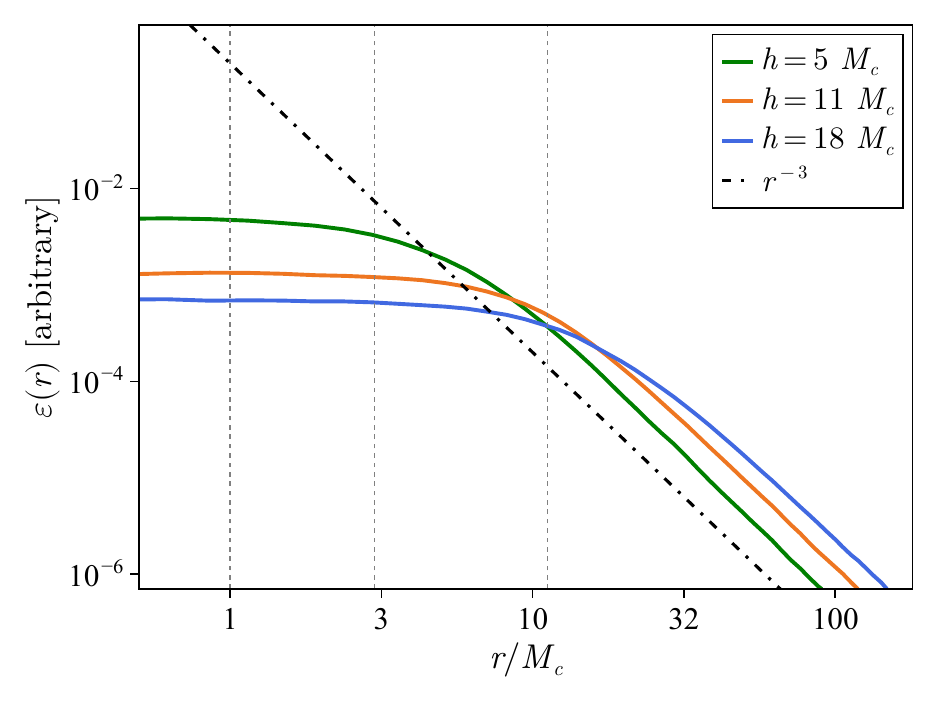}
    \caption{Radial emissivity profiles for various lamp-post corona heights $h$, computed within a RAR-DM background of $0.7\Ucr$ compactness. Vertical dash lines show different inner radii for the accretion disk $\Rin=\Rsat, 3M_c, r_c$. }
    \label{fig:emissivities}
\end{figure}

The aim of this section is to compare morphologically the resulting iron line profiles in a RAR setup to those in the black hole scenario. As is standard for relativistic line modeling \citep{Bambi2013b,Dauser2013}, all profiles were normalized to unit area to facilitate direct comparison across a consistent scale. Therefore, a higher line peak does not imply a higher energy flux.
In this subsection we compute lines corresponding to a lamp-post corona prescription at a fixed height of $h=5M$ and photon index $\Gamma=2$. 

In Fig. \ref{fig:mosaic_1} we show a mosaic-type plot where the rows correspond to emission lines for different compactness. 
In the columns we vary on the viewing inclination angle, increasing from left to right. 
In all subplots we show for comparison lines generated with black hole metric with extreme spin parameters, $a=0$ in solid gray, and $a=1$ in dash-doted black (the disks start at the corresponding ISCO radii, $\Rin=6\MBH$ and $\Rin=1\MBH$). 
In the RAR case, we include two possibilities for the inner radius of the accretion disk. 
In solid colored line, the disk starts at the corresponding value of the core for the given compactness, $\Rin=r_c$ (see table \ref{tab:table_1}). 
In colored dash line, the disk starts at $\Rin=6M_c$, to compare with the Schwarzschild black hole (in solid gray).
As the compactness decreases, the red-wing onset of the emission line shifts toward higher energies, resulting in more skewed profiles regardless of the inclination angle. 
This trend is not solely driven by the increase in the core radius $r_c$ as compactness lowers; it also arises from inherent properties of the background metric that become more pronounced with higher compactness. 
A consistent behavior is observed for the dashed lines, which represent disks extending from $\Rin=6M_c$.
The transition to a double-horn structure at lower compactness (last row), even with emission starting at $r_c$, serves as a distinct signature of the diminished influence of strong-field gravity at these scales, where Doppler broadening begins to dominate over relativistic effects.

In Fig. \ref{fig:retrograde_spin} we show the similarities between Fe K$\alpha$ emission lines from accretion disks in retrograde orbits around Kerr black holes (with spin parameters $-1 < a < 0$) and those around fermionic dark matter cores.
We find that by decreasing the compactness of the DM core, it is possible to reproduce the full range of retrograde spin parameter, spanning from $\Ucr$ down to $0.7\Ucr$. 

In Fig. \ref{fig:E_min_comparison} we show the trends of the minimum energy achieved for Kerr solutions (i.e. the boundary of the red-wing) as a function of spin parameter, and for RAR solutions as a function of fractional compactness. These values correspond to fixed inclination angle of $i=45^\circ$. 
We can see here that only the most compact solutions can produce broadened lines such as those observed for black holes, but for spin values $a<0$ as shown in the shaded region. 
This result is similar to those found for other regular spacetimes \citep[e.g.,][]{Bambi2013a}. Even though for (most) regular solutions, there is no ISCO and the accretion disk can extend up to much more smaller radii, the iron line does not present the extended red-wing observed for rapidly rotating BHs.

\begin{figure*}[t]
    \centering
    \includegraphics[width=0.9\linewidth]{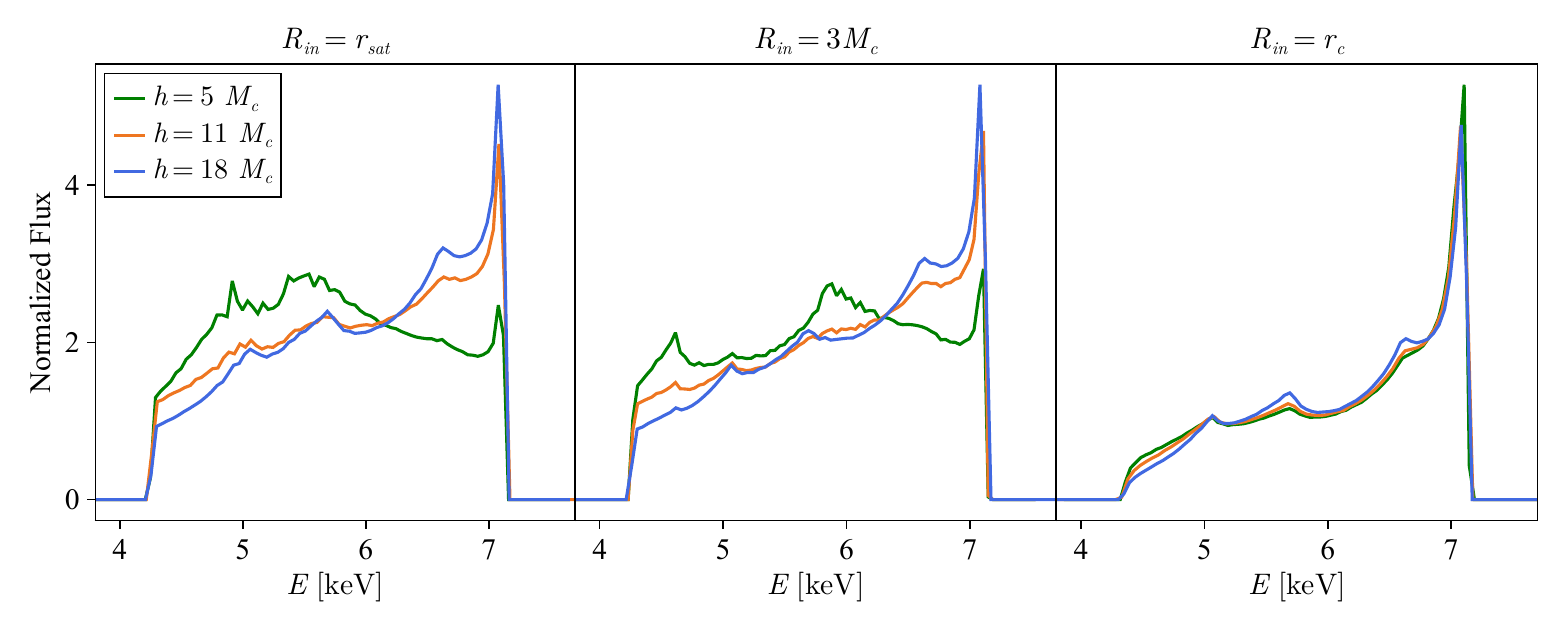}
    \caption{Fe K$\alpha$ line profiles within the lamp-post corona prescription as we vary the corona height $h$, for the same background of $0.7\Ucr$ and same inclination angle $i=45^\circ$. We show different values of the inner radius $\Rin$ in the three panels. The emissivity profiles correspond to the ones in Fig. \ref{fig:emissivities}.}
    \label{fig:triple-heights}
\end{figure*}

\begin{figure*}[t]
    \centering
    \includegraphics[width=0.9\linewidth]{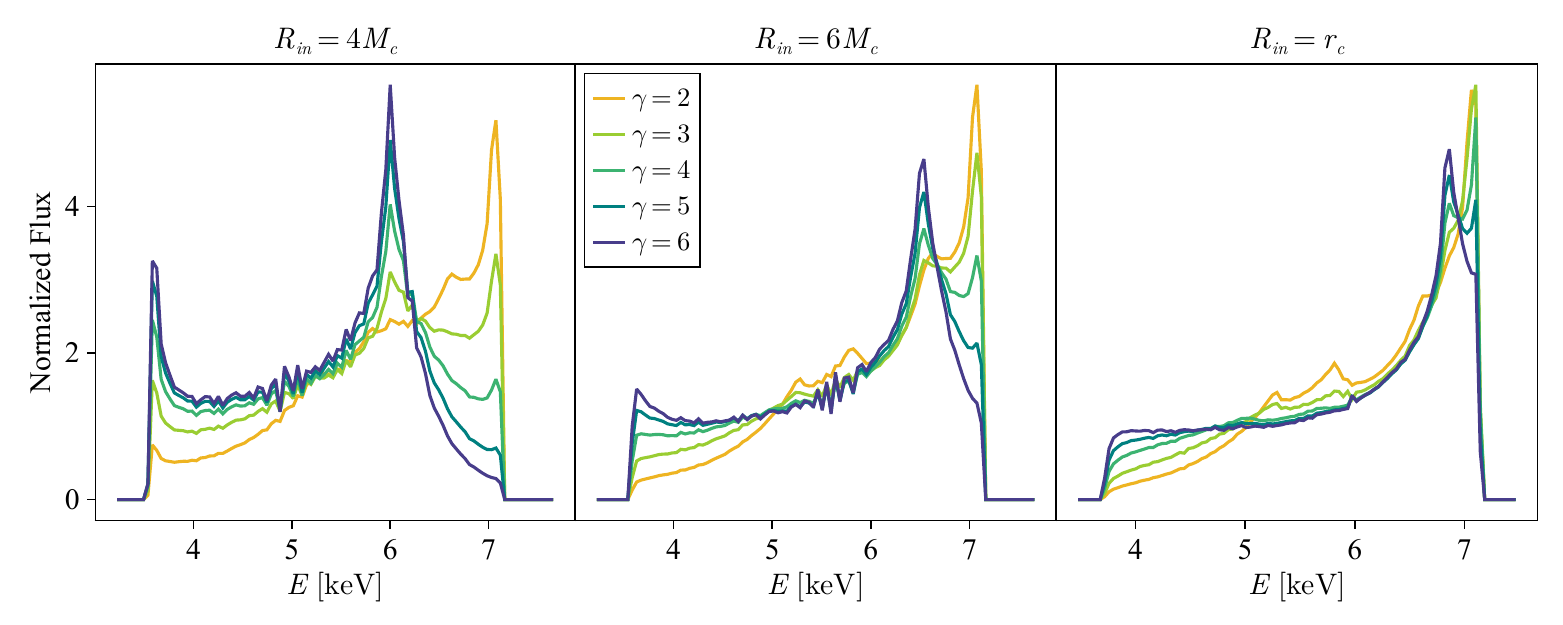}
    \caption{Fe K$\alpha$ line profiles for disks with power-law emissivity functions of index $\gamma$. These lines are computed for $i=45^\circ$ inclination angle under a background metric given by the most compact regular solution $\Ucr$. In the different panels we change the onset of the accretion disk, given by $\Rin$.}
    \label{fig:PL-378-1}    
\end{figure*}

\subsection{Different disk-corona configurations}

In this section we further explore the parameters of the disk-corona configurations. 
Fig. \ref{fig:emissivities} shows the radial emissivity profiles on the accretion disk for different heights of the lamp-post corona model. 
Consistently, as the corona source is positioned closer to the center, a higher fraction of emitted photons is focused on the inner parts of the disk, where the emissivity remains approximately constant. 
Conversely, at larger radii, the radial profile exhibits a characteristic $r^{-3}$ behavior, typical of a flat spacetime regime. Vertical dotted lines mark the selected inner disk radii, $\Rin=\Rsat, 3M_c, r_c$. (see table \ref{tab:table_1}). 
In Fig. \ref{fig:triple-heights}, we show the corresponding iron line profiles for different lamp-post corona heights and disk radii, using the same emissivity profiles.
As shown in the right panel, the iron line profiles for disks extending down to the core radius $\Rin=r_c$, are almost independent of the corona height, since the corresponding emissivity functions remain very similar from $r \geq r_c$ in Fig. \ref{fig:emissivities}.

Interestingly, a different morphology appears for disks extending well inside the core and for low corona heights, as shown by the green line in the left and center panels of Fig. \ref{fig:triple-heights}. 
In these conditions, the emission is predominantly weighted toward the innermost regions of the disk, where the orbital velocities of the emitting particles experience a significant decrease, dropping by nearly two orders of magnitude compared with particles located at the core radius.
Consequently, this kinematic suppression directly alters the line morphology, yielding line profiles broad but notably flattened, even blurring the characteristic blue peak. 
Fig. \ref{fig:triple-heights} only considers a background metric corresponding to the $0.7\Ucr$ configuration; nevertheless, the same features are obtained for other compactness.

This behavior contrasts with the seminal work by \citet{Reynolds1997}, who studied the contribution of reflection emission from the plunging region inside the ISCO of a Schwarzschild black hole. They showed that, provided the X-ray source is inefficient enough not to over-ionize the in-falling material, emission from this region can significantly  contribute to the observed iron line, producing an asymmetric profile characterized by a prominent blue peak and an extended red-wing for low inclination angles \citep[see Fig. 4 in][]{Reynolds1997}. 
Conversely, although the RAR accretion disk extends much closer to the center, the shallower gravitational potential associated with the RAR distribution modifies both the orbital velocity profile and the gravitational redshift experienced by the emitting plasma. As a result, the flattened line profiles are only observed for lower corona heights, while they disappear as the corona gets higher.

\begin{figure}[t]
    \centering
    \includegraphics[width=0.9\columnwidth]{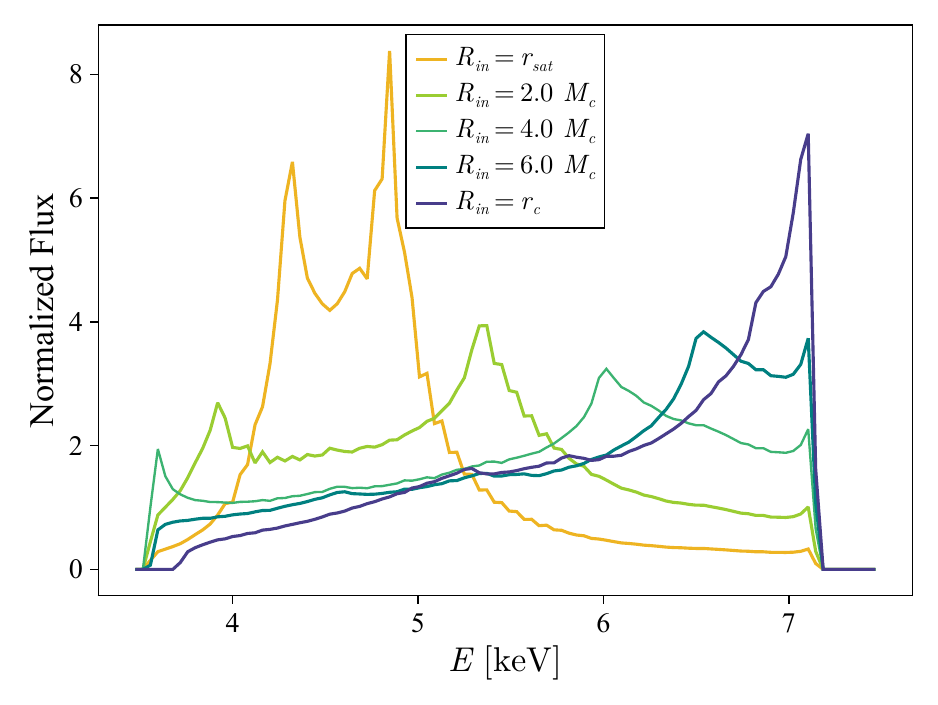}
        \caption{Fe K$\alpha$ line profiles obtained for a disk intensity $I(r) \propto r^{-3}$, as a function of the inner radius of the accretion disk. All cases are for $i=45^\circ$, and for the critical compactness of the DM core $\Ucr$.}
    \label{fig:PL-378-2}
\end{figure}

To further compare with previous works, we also compute iron line profiles using the phenomenological power-law emissivity prescription given by Eq. \ref{eq:intensity-PL}.
These results are shown in Figs. \ref{fig:PL-378-1} and \ref{fig:PL-378-2}. For these calculations, we adopt the background metric given by $\Ucr$ and fix the inclination angle to $i=45^\circ$ for all profiles. 

In Fig. \ref{fig:PL-378-1}, we vary the power-law index $\gamma$ from $2$ to $6$, for the three values considered for the inner disk radius $\Rin$. Under this prescription, all disks show the same qualitative trend: increasing the power-law index enhances the contribution from the innermost disk regions, producing a stronger red-wing and shifting the dominant blue peak toward lower energies. However, varying this parameter does not significantly increase the width of the emission line.

In Fig. \ref{fig:PL-378-2}, we set the power-law index to $\gamma=3$, and vary on the inner radius of the accretion disk. In this case, we recover the behavior reported in \cite[][fig. 5]{Bambi2013b} for the regular perfect-fluid case. 

\section{A case study: MCG-06-30-15} \label{sec:mcg}

MCG-06-30-15 is a nearby narrow-line Seyfert 1 galaxy located at $37$ Mpc ($z=0.0077$, \citet{Pineda1980}). It was the first AGN in which a broad Fe K$\alpha$ line was observed \citep{Tanaka1995,Iwasawa1996}, and since then, it has become the archetypal source for studies of relativistic reflection in AGN. 
The source is highly variable, and it is detected in different spectral states, similar to those observed in X-ray binaries \citep{Fabian2003}. 
Its X-ray spectrum shows strong reflection features, indicating emission from regions very close to the central compact object. 
MCG-06-30-15 has been extensively observed by multiple X-ray observatories, and these studies consistently support the presence of broad and strongly redshifted Fe K$\alpha$ line from the innermost regions of an accretion disk \citep[e.g.,][]{Nandra1989,Lee1998,Wilms2001,Fabian2002,Young2005,Brenneman2006,Miniutti2007,Marinucci2014}. 

The X-ray continuum is well described by emission from a hot, compact corona, modeled as a power-law, together with a reflection component produced in the inner accretion disk illuminated by the corona. Additional components are usually needed to reproduce both soft and hard X-rays, including a warm absorber that modifies the spectra below $\sim 3$ keV, and the reflection bump which affects the continuum mainly above $\sim 8$ keV. 

The SMBH candidate at the center of MCG-06-30-15 is estimated to have a mass of $(3.6-6) \times 10^{6} M_{\odot}$ \citep{McHardy2005}.
Measurements based on relativistic reflection modeling of the iron line indicate, under the black hole paradigm, high rotation: early estimates gave a dimensionless spin parameter of $a\sim 0.989$ \citep{Fabian2002,Brenneman2006}, while more recent studies show $a>0.65$ \citep{Brenneman2025}. 
Recent spectral modeling with data from {X-Ray Imaging and Spectroscopy Mission} (XRISM), also supports a high spin parameter (first with a time average analysis \citep{Brenneman2025} and later with a time resolved analysis \citep{Wilkins2026}, implying that the accretion disk extends very close to the central object. These conditions make MCG–06-30-15 a particularly probe of the strong-gravity regime.

\begin{table}
    \caption{Simple continuum model fit inferred with the \textsc{xspec} tool.}
    \centering
    \begin{tabular}{|l|l|c|} \hline
       Component  & Parameter & Value \\ \hline \hline
       \texttt{tbabs} & $N_H\ [10^{22}]$ & $0.304 \pm 0.099$ \\ \hline
       \texttt{powerlaw} & $\Gamma$ & $1.995 \pm 0.026$ \\
       \texttt{powerlaw} & $K\ [10^{-2}]$ & $1.276 \pm 0.065$ \\ \hline
    \end{tabular}
    \label{tab:cont_model}
\end{table}

\begin{table}
    \caption{Fe K$\alpha$ line fits corresponding to Fig. \ref{fig:MCG_LP} for the lamp-post corona prescription, and Fig. \ref{fig:MCG_PL} for the phenomenological power-law emissivity prescription.}
    \centering
    \begin{tabular}{|l|l||c|c|c|c|} \hline
        \multicolumn{6}{|c|}{Lamp-post Corona} \\ \hline \hline
        $M/R$ & line & $h\ [M_c]$ & $\Gamma$ & $\Rin\ [M_c]$ & $i$ [°] \\ \hline
        $\Ucr$ & {\small red solid} & $3.0$ & $2.0$ & $r_c$ & $35$ \\ 
        $\Ucr$ & {\small red dash} & $5.0$ & $2.0$ & $6.0$ & $35$ \\ \hline
        $0.7\Ucr$ & {\small green solid} & $5.0$ & $2.0$ & $8.0$ & $32$ \\ 
        $0.7\Ucr$ & {\small green dash} & $18.0$ & $2.0$ & $r_c$ & $29$ \\ \hline \hline 
        \multicolumn{6}{|c|}{Power Law} \\ \hline \hline
        \multicolumn{1}{|l|}{$M/R$} & \multicolumn{1}{l||}{line} & \multicolumn{2}{c|}{$\gamma$} & \multicolumn{1}{c|}{$\Rin\ [M_c]$} & \multicolumn{1}{c|}{$i$ [°]} \\ \hline
        $\Ucr$ & {\small pink solid} & \multicolumn{2}{c|}{$6.0$} & $6.0$ & $33$ \\ 
        $\Ucr$ & {\small pink dash} & \multicolumn{2}{c|}{$6.0$} & $r_c$ & $30$ \\ \hline
        $0.7\Ucr$ & {\small blue solid} & \multicolumn{2}{c|}{$4.0$} & $9.0$ & $44$ \\ 
        $0.7\Ucr$ & {\small blue dash} & \multicolumn{2}{c|}{$5.0$} & $r_c$ & $37$ \\ \hline 
    \end{tabular}
    \label{tab:table_3}
\end{table}

\begin{figure}[t]
    \centering
    \includegraphics[width=0.9\columnwidth]{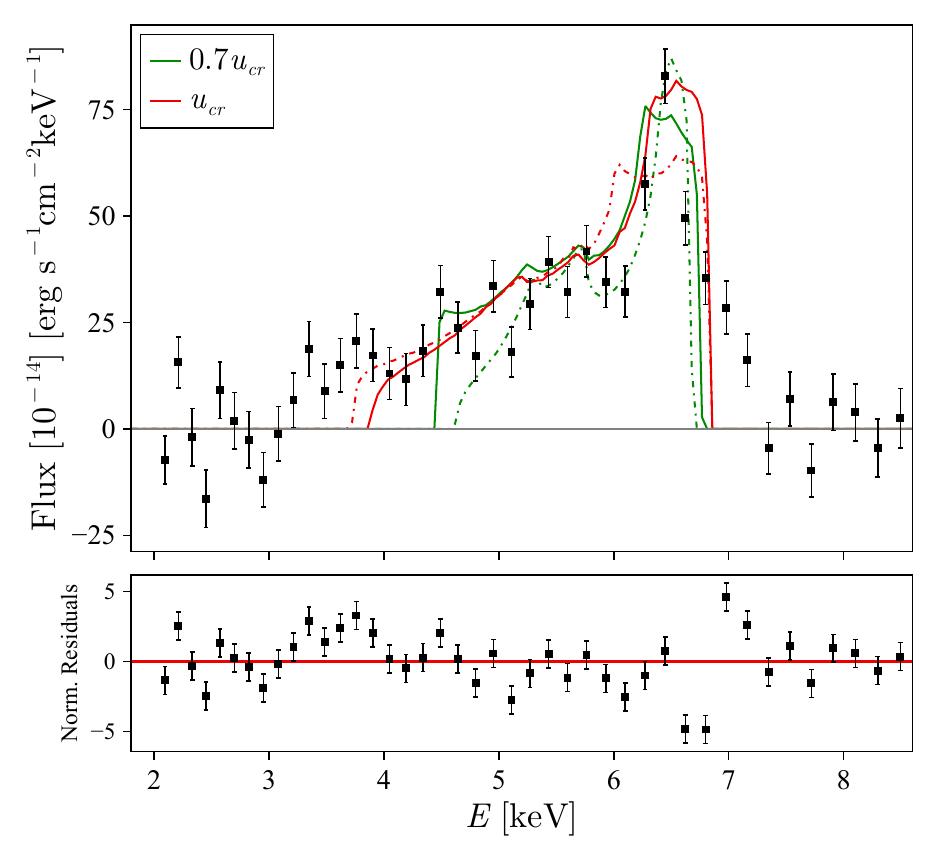}
    \caption{Line emission fits in the lamp-post corona prescription. Red lines correspond to a background generated by the critic compactness configuration $\Ucr$, and green lines correspond to $0.7\Ucr$. We vary on the geometrical parameters $i$, $h$ and $\Rin$ to adjust the line. Values obtained are reported in table \ref{tab:table_3}. The bottom panel shows the normalized residuals with respect to the solid red line model.}
    \label{fig:MCG_LP}
\end{figure}

\begin{figure}[t]
    \centering
    \includegraphics[width=0.9\columnwidth]{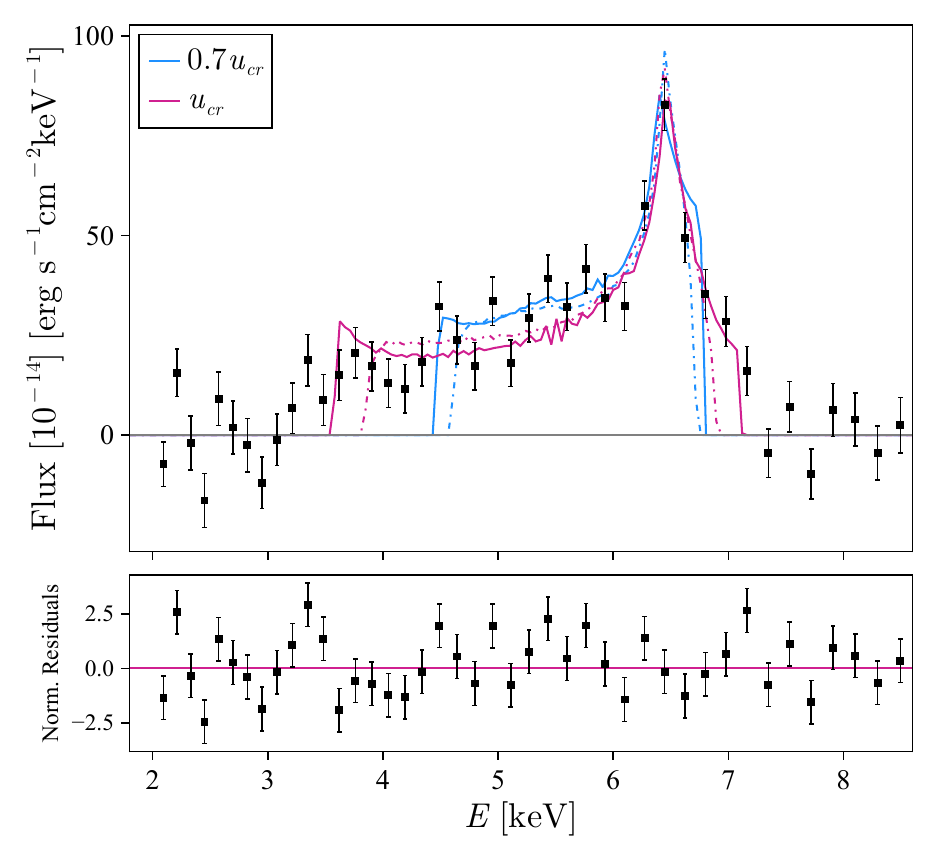}
    \caption{Line emission fits in the phenomenological Power Law prescription. Pink lines correspond to a background generated by the critic compactness configuration $\Ucr$, and blue lines correspond to $0.7\Ucr$. We vary on the geometrical parameters $i$, $\gamma$ and $\Rin$ to adjust the line. Values obtained are reported in table \ref{tab:table_3}.The bottom panel shows the normalized residuals with respect to the solid pink line model.}
    \label{fig:MCG_PL}
\end{figure}

In this section, we aim to assess the viability of the predicted Fe K$\alpha$ lines under a RAR-DM background to reproduce real observations of a relativistically broadened emission line. 
To do so, we have reduced an observation (Obs.ID 0693781301) of MCG-06-30-15 obtained with the XMM-Newton observatory during revolution 2408 in 2013 \citep{Marinucci2014}. 
The data were processed following standard reduction procedures with the Science Analysis System software (SAS, v22.1) and calibrated using the latest Current Calibration Files (CCF).
We have only used data from the EPIC-pn camera, which operated in the small-window mode. After filtering for high background episodes, we obtained a clean GTI file with net exposure time of $\sim 91$ ks. 
We selected circular extraction regions of $20''$ radius for both source and background each. We discarded photon pile-up distortion by verifying the event photon statistics with the \texttt{epatplot} task.
The instrumental response was accounted for by generating the respective Redistribution Matrix Files (RMF) and Ancillary Response Files (ARF).

We modeled the underlying continuum using the \textsc{xspec} tool \citep{Arnaud1996}\footnote{Spectral fitting package from \href{https://heasarc.gsfc.nasa.gov/docs/software/xspec/index.html}{HEASARC.}}, excluding the broad line band by fitting only on the ranges $(2.0,3.5)$ keV and $(7.5,10.5)$ keV. 
We assumed a primary power-law continuum modified by cold absorption from the interstellar medium \texttt{(tbabs*powerlaw)}, given that 
any reflection component is expected to be negligible across these specific bands.
The corresponding components and their best fit values are listed in table \ref{tab:cont_model}.

In Fig.s \ref{fig:MCG_LP} and \ref{fig:MCG_PL} we show the data line profile after continuum subtraction, together with modeled lines within the extended RAR-DM background with different compactness values. 
We selected the optimal line profiles when we vary different physical and geometrical parameters. 
Fig. \ref{fig:MCG_LP} corresponds to accretion disks illuminated under the lamp-post corona prescription. 
There, we show configurations for the critic compactness solution $\Ucr$ in red lines and $0.7\Ucr$ in green lines, for different corona heights $h$, inner radius $\Rin$ and inclination angles $i^\circ$. 
The photon index $\Gamma$ adopts the same value as the fitted continuum. Bottom panel shows the normalized residuals to the (visually) optimal line. 
Fig. \ref{fig:MCG_PL} shows the equivalent as the previous figure, but for accretion disks emissivity profiles under a power-law prescription. 
Configurations for the critic compactness solution $\Ucr$ are shown in pink lines and $0.7\Ucr$ in blue lines, for different indexes $\gamma$, inner radius $\Rin$ and inclination angles $i^\circ$. 
All parameter fit values are reported in table \ref{tab:table_3}.

Overall, the predicted profiles achieve a reasonable agreement with the observed broad features, provided that near critical compactness configurations are considered. 
We note, however, that while the observations includes the entire iron K-complex (i.e. including potential Fe K$\beta$ emission and various ionization states) our model accounts solely for the relativistic Fe K$\alpha$ line morphology. This simplification introduces minor residual features near $\sim 7$~keV that remain unaccounted for.

Furthermore, we note that while we have used historical XMM-Newton datasets for the spectral analysis of MCG-06-30-15, very recent observations from the X-Ray Imaging and Spectroscopy Mission (XRISM) have provided unprecedented high-resolution views of this source \citep{Brenneman2025}. Incorporating the new XRISM data lies beyond the scope of this initial phenomenological paper, as it requires a dedicated, computationally intensive modeling of the complex warm absorber components and micro-calorimeter resolution lines. Nevertheless, XRISM data strongly support the presence of the inner-disk reflection features over multizone-absorbers models \citep{Young2005,Marinucci2014}, which validate our approach.

\section{Discussion and conclusions} \label{sec:conclusions}

In this work, we have investigated the relativistic iron K$\alpha$ line profiles produced by accretion disks around regular fermionic dark matter cores described within the extended Ruffini–Argüelles–Rueda (RAR) model. 
Using the general relativistic ray-tracing code \texttt{Skylight}, we explored how the core compactness $u$ --determined by the DM fermion mass $m_f$-- and the inner disk radius, affect the resulting spectral signatures. 
Additionally, we applied these models to observations of the Seyfert 1 galaxy MCG-06-30-15 to test the ability of RAR configurations to reproduce features commonly associated with SMBHs.

Our analysis highlights several distinctive features of iron-line emission in non-singular compact objects. First, the relative position of the lamp-post corona and the fermionic core plays a crucial role in determining the disk illumination profile. For corona heights larger than the core radius ($h>r_c$), the emissivity profile approaches the standard asymptotic behavior expected for extended disks. Conversely, when the corona is located inside the core ($h<r_c$), the shallower gravitational potential modifies the illumination of the innermost disk regions. 

The absence of an ISCO in the RAR spacetime allows stable circular motion at radii smaller than the core radius, raising the question of the physical extent of the accretion disk. We find that disks truncated near the core radius ($\Rin=r_c$) produce iron-line profiles that are largely insensitive to variations in the corona height. However, when the disk extends deeper into the core, down to the binding-energy saturation radius ($\Rsat$), and the corona is located at low heights, the line morphology changes significantly. In this regime, the reduced orbital velocities inside the core produces broad yet flattened profiles in which the characteristic blue peak becomes strongly weakened.

We also find that the most compact RAR configurations can reproduce several characteristic features of black hole iron lines, including broad red-wings and prominent blue horns. Nevertheless, they do not fully match the extreme red-wing extensions as theoretically predicted by rapidly rotating prograde Kerr black holes ($a \gtrsim 0.9$). 

When applied to X-ray observations of MCG-06-30-15, the fermionic DM core provides a viable alternative interpretation of the observed broad iron line to that associated with a Kerr black hole. Although standard reflection models generally associate this source with a rapidly spinning SMBH, the critical RAR configuration ($\Ucr =0.128$) can reproduce the main spectral features without invoking an event horizon or a central spin parameter. This degeneracy illustrates that iron-line spectroscopy alone may not uniquely identify the nature of the central compact object (see also \citep{2021SSRv..217...65B} for similar conclusions).

The observed reflection features are highly dependent on the spacetime geometry in the innermost disk regions. 
In this context, alternative compact object models can produce observational signatures that are difficult to distinguish from those expected from a Kerr black hole. 
For instance, fermionic core-halo solutions, despite being non-rotating, allows stable circular orbits down to very small radii, potentially reproducing key features typically attributed to black hole spin. 
In addition, \citet{Patel2026} showed that the hard X-ray spectrum can also be reproduced by an accretion disk around a non-spinning naked singularity. These results show that an independent measurement of the spin of the compact object is fundamental for breaking the degeneracies between different models. 

It should be noted that the phenomenological analysis of the broadened $\text{Fe K}\alpha$ line profile in MCG-6-30-15 presented in this work is intended as an initial exploration. To further complement and validate these results, a broader spectral analysis spanning a wider energy range (from approximately $0.1$ to $50\text{ keV}$) should be performed. The extended spectra would allow for a more detailed assessment of the complete reflection component and a more precise modeling of the complex absorption features known to characterize this source \citep{Miniutti2007,Brenneman2025}, both of which can subtly influence the profile of the prominent red-wing. 
Additionally, currently available X-ray spectral modeling are suited for a standard Kerr spacetime framework. 
A detailed treatment of relativistic reflection processes in the specific RAR-DM background is left for future work, as it goes beyond the scope of this introductory paper.

In conclusion, while iron-line spectroscopy remains a powerful probe of strong-field gravity, additional independent constraints are required to distinguish SMBHs from alternative compact objects such as fermionic dark matter cores.

\begin{acknowledgements}

We thank K. Iwasawa for his assistance with data reduction and J. Pelle for his guidance with the Skylight package.\\
This work used computational resources from CCAD – Universidad Nacional de Córdoba (\href{https://ccad.unc.edu.ar/}{https://ccad.unc.edu.ar/}), which are part of SNCAD – MinCyT, República Argentina. \\
V.C. and C.R.A thanks financial support from CONICET, Argentina. \\
We thank IALP support and administrative staff for their dedicated work. \\

\end{acknowledgements}

%
\bibliographystyle{aa_url} 
\bibliography{references} 

@ARTICLE{Boltzmann1884,
       author = {{Boltzmann}, Ludwig},
        title = "{Ableitung des Stefan'schen Gesetzes, betreffend die Abh{\"a}ngigkeit der W{\"a}rmestrahlung von der Temperatur aus der electromagnetischen Lichttheorie}",
      journal = {Annalen der Physik},
         year = 1884,
        month = jan,
       volume = {258},
       number = {6},
        pages = {291-294},
          doi = {10.1002/andp.18842580616},
       adsurl = {https://ui.adsabs.harvard.edu/abs/1884AnP...258..291B}
}

@ARTICLE{Tolman1930,
       author = {{Tolman}, Richard C.},
        title = "{On the Weight of Heat and Thermal Equilibrium in General Relativity}",
      journal = {Physical Review},
         year = 1930,
        month = apr,
       volume = {35},
       number = {8},
        pages = {904-924},
          doi = {10.1103/PhysRev.35.904},
       adsurl = {https://ui.adsabs.harvard.edu/abs/1930PhRv...35..904T}
}

@ARTICLE{Oppenheimer1939,
       author = {{Oppenheimer}, J.~R. and {Volkoff}, G.~M.},
        title = "{On Massive Neutron Cores}",
      journal = {Physical Review},
         year = 1939,
        month = feb,
       volume = {55},
       number = {4},
        pages = {374-381},
          doi = {10.1103/PhysRev.55.374},
       adsurl = {https://ui.adsabs.harvard.edu/abs/1939PhRv...55..374O}
}

@ARTICLE{Buchdal1959,
       author = {{Buchdahl}, H.~A.},
        title = "{General Relativistic Fluid Spheres}",
      journal = {Physical Review},
         year = 1959,
        month = nov,
       volume = {116},
       number = {4},
        pages = {1027-1034},
          doi = {10.1103/PhysRev.116.1027},
       adsurl = {https://ui.adsabs.harvard.edu/abs/1959PhRv..116.1027B}
}

@BOOK{Harrison1965,
       author = {{Harrison}, B.~K. and {Thorne}, K.~S. and {Wakano}, M. and {Wheeler}, J.~A.},
        title = "{Gravitation Theory and Gravitational Collapse}",
         year = 1965,
         publisher={University of Chicago Press},
       adsurl = {https://ui.adsabs.harvard.edu/abs/1965gtgc.book.....H}
}

@ARTICLE{Lindquist1966,
       author = {{Lindquist}, Richard W.},
        title = "{Relativistic transport theory}",
      journal = {Annals of Physics},
         year = 1966,
        month = may,
       volume = {37},
       number = {3},
        pages = {487-518},
          doi = {10.1016/0003-4916(66)90207-7},
       adsurl = {https://ui.adsabs.harvard.edu/abs/1966AnPhy..37..487L}
}

@ARTICLE{Ruffini1969,
       author = {{Ruffini}, Remo and {Bonazzola}, Silvano},
        title = "{Systems of Self-Gravitating Particles in General Relativity and the Concept of an Equation of State}",
      journal = {Physical Review},
         year = 1969,
        month = nov,
       volume = {187},
       number = {5},
        pages = {1767-1783},
          doi = {10.1103/PhysRev.187.1767},
       adsurl = {https://ui.adsabs.harvard.edu/abs/1969PhRv..187.1767R}
}

@book{Weinberg1972,
    author = "Weinberg, Steven",
    title = "{Gravitation and Cosmology}: {Principles and Applications of the General Theory of Relativity}",
    isbn = "978-0-471-92567-5, 978-0-471-92567-5",
    publisher = "John Wiley and Sons",
    address = "New York",
    year = "1972"
}

@ARTICLE{Shakura1973,
       author = {{Shakura}, N.~I. and {Sunyaev}, R.~A.},
        title = "{Black holes in binary systems. Observational appearance.}",
      journal = {\aap},
         year = 1973,
        month = jan,
       volume = {24},
        pages = {337-355},
       adsurl = {https://ui.adsabs.harvard.edu/abs/1973A&A....24..337S}
}

@ARTICLE{Pineda1980,
       author = {{Pineda}, F.~J. and {Delvaille}, J.~P. and {Grindlay}, J.~E. and {Schnopper}, H.~W.},
        title = "{X-ray and optical observations of MCG -6-30-15.}",
      journal = {\apj},
         year = 1980,
        month = apr,
       volume = {237},
        pages = {414-417},
          doi = {10.1086/157883},
       adsurl = {https://ui.adsabs.harvard.edu/abs/1980ApJ...237..414P}
}

@ARTICLE{Sunyaev1980,
       author = {{Sunyaev}, R.~A. and {Titarchuk}, L.~G.},
        title = "{Comptonization of X-Rays in Plasma Clouds - Typical Radiation Spectra}",
      journal = {\aap},
         year = 1980,
        month = jun,
       volume = {86},
        pages = {121},
       adsurl = {https://ui.adsabs.harvard.edu/abs/1980A&A....86..121S}
}

@ARTICLE{Sorkin1981,
       author = {{Sorkin}, R.},
        title = "{A Criterion for the Onset of Instability at a Turning Point}",
      journal = {\apj},
         year = 1981,
        month = oct,
       volume = {249},
        pages = {254},
          doi = {10.1086/159282},
       adsurl = {https://ui.adsabs.harvard.edu/abs/1981ApJ...249..254S}
}

@BOOK{Shapiro1983,
       author = {{Shapiro}, Stuart L. and {Teukolsky}, Saul A.},
        title = "{Black holes, white dwarfs and neutron stars. The physics of compact objects}",
         year = 1983,
         publisher = "John Wiley and Sons",
          doi = {10.1002/9783527617661},
       adsurl = {https://ui.adsabs.harvard.edu/abs/1983bhwd.book.....S}
}

@ARTICLE{Nandra1989,
       author = {{Nandra}, K. and {Pounds}, K.~A. and {Stewart}, G.~C. and {Fabian}, A.~C. and {Rees}, M.~J.},
        title = "{Detection of iron features in the X-ray spectrum of the Seyfert I galaxy MCG -6-30-15.}",
      journal = {\mnras},
         year = 1989,
        month = jan,
       volume = {236},
        pages = {39P-46},
          doi = {10.1093/mnras/236.1.39P},
       adsurl = {https://ui.adsabs.harvard.edu/abs/1989MNRAS.236P..39N}
}

@ARTICLE{Fabian1989,
       author = {{Fabian}, A.~C. and {Rees}, M.~J. and {Stella}, L. and {White}, N.~E.},
        title = "{X-ray fluorescence from the inner disc in Cygnus X-1.}",
      journal = {\mnras},
         year = 1989,
        month = may,
       volume = {238},
        pages = {729-736},
          doi = {10.1093/mnras/238.3.729},
       adsurl = {https://ui.adsabs.harvard.edu/abs/1989MNRAS.238..729F}
}

@ARTICLE{Laor1991,
       author = {{Laor}, Ari},
        title = "{Line Profiles from a Disk around a Rotating Black Hole}",
      journal = {\apj},
         year = 1991,
        month = jul,
       volume = {376},
        pages = {90},
          doi = {10.1086/170257},
       adsurl = {https://ui.adsabs.harvard.edu/abs/1991ApJ...376...90L}
}

@ARTICLE{Haardt1991,
       author = {{Haardt}, F. and {Maraschi}, L.},
        title = "{A Two-Phase Model for the X-Ray Emission from Seyfert Galaxies}",
      journal = {\apjl},
         year = 1991,
        month = oct,
       volume = {380},
        pages = {L51},
          doi = {10.1086/186171},
       adsurl = {https://ui.adsabs.harvard.edu/abs/1991ApJ...380L..51H}
}

@ARTICLE{George1991,
       author = {{George}, I.~M. and {Fabian}, A.~C.},
        title = "{X-ray reflection from cold matter in Active Galactic Nuclei and X-ray binaries.}",
      journal = {\mnras},
         year = 1991,
        month = mar,
       volume = {249},
        pages = {352},
          doi = {10.1093/mnras/249.2.352},
       adsurl = {https://ui.adsabs.harvard.edu/abs/1991MNRAS.249..352G}
}

@ARTICLE{Tanaka1995,
       author = {{Tanaka}, Y. and {Nandra}, K. and {Fabian}, A.~C. and {Inoue}, H. and {Otani}, C. and {Dotani}, T. and {Hayashida}, K. and {Iwasawa}, K. and {Kii}, T. and {Kunieda}, H. and {Makino}, F. and {Matsuoka}, M.},
        title = "{Gravitationally redshifted emission implying an accretion disk and massive black hole in the active galaxy MCG-6-30-15}",
      journal = {\nat},
         year = 1995,
        month = jun,
       volume = {375},
       number = {6533},
        pages = {659-661},
          doi = {10.1038/375659a0},
       adsurl = {https://ui.adsabs.harvard.edu/abs/1995Natur.375..659T}
}

@ARTICLE{Martocchia1996,
       author = {{Martocchia}, Andrea and {Matt}, Giorgio},
        title = "{Iron Kalpha line intensity from accretion discs around rotating black holes}",
      journal = {\mnras},
         year = 1996,
        month = oct,
       volume = {282},
       number = {4},
        pages = {L53-L57},
          doi = {10.1093/mnras/282.4.L53},
       adsurl = {https://ui.adsabs.harvard.edu/abs/1996MNRAS.282L..53M}
}

@ARTICLE{Iwasawa1996,
       author = {{Iwasawa}, K. and {Fabian}, A.~C. and {Reynolds}, C.~S. and {Nandra}, K. and {Otani}, C. and {Inoue}, H. and {Hayashida}, K. and {Brandt}, W.~N. and {Dotani}, T. and {Kunieda}, H. and {Matsuoka}, M. and {Tanaka}, Y.},
        title = "{The variable iron K emission line in MCG-6-30-15}",
      journal = {\mnras},
         year = 1996,
        month = oct,
       volume = {282},
       number = {3},
        pages = {1038-1048},
          doi = {10.1093/mnras/282.3.1038},
archivePrefix = {arXiv},
       eprint = {astro-ph/9606103},
 primaryClass = {astro-ph},
       adsurl = {https://ui.adsabs.harvard.edu/abs/1996MNRAS.282.1038I}
}

@INPROCEEDINGS{Arnaud1996,
       author = {{Arnaud}, K.~A.},
        title = "{XSPEC: The First Ten Years}",
    booktitle = {Astronomical Data Analysis Software and Systems V},
         year = 1996,
       editor = {{Jacoby}, George H. and {Barnes}, Jeannette},
       series = {Astronomical Society of the Pacific Conference Series},
       volume = {101},
        month = jan,
        pages = {17},
       adsurl = {https://ui.adsabs.harvard.edu/abs/1996ASPC..101...17A}
}

@ARTICLE{Matt1996,
       author = {{Matt}, G. and {Fabian}, A.~C. and {Ross}, R.~R.},
        title = "{Iron K fluorescent lines from relativistic, ionized discs}",
      journal = {\mnras},
         year = 1996,
        month = feb,
       volume = {278},
       number = {4},
        pages = {1111-1120},
          doi = {10.1093/mnras/278.4.1111},
       adsurl = {https://ui.adsabs.harvard.edu/abs/1996MNRAS.278.1111M}
}

@ARTICLE{Henri1997,
       author = {{Henri}, G. and {Petrucci}, P.~O.},
        title = "{Anisotropic illumination of AGN's accretion disk by a non thermal source. I. General theory and application to the Newtonian geometry.}",
      journal = {\aap},
         year = 1997,
        month = oct,
       volume = {326},
        pages = {87-98},
          doi = {10.48550/arXiv.astro-ph/9705233},
archivePrefix = {arXiv},
       eprint = {astro-ph/9705233},
 primaryClass = {astro-ph},
       adsurl = {https://ui.adsabs.harvard.edu/abs/1997A&A...326...87H}
}

@ARTICLE{Fanton1997,
       author = {{Fanton}, Claudio and {Calvani}, Massimo and {de Felice}, Fernando and {Cadez}, Andrej},
        title = "{Detecting Accretion Disks in Active Galactic Nuclei}",
      journal = {\pasj},
         year = 1997,
        month = apr,
       volume = {49},
        pages = {159-169},
          doi = {10.1093/pasj/49.2.159},
       adsurl = {https://ui.adsabs.harvard.edu/abs/1997PASJ...49..159F}
}

@ARTICLE{Reynolds1997,
       author = {{Reynolds}, Christopher S. and {Begelman}, Mitchell C.},
        title = "{Iron Fluorescence from within the Innermost Stable Orbit of Black Hole Accretion Disks}",
      journal = {\apj},
         year = 1997,
        month = oct,
       volume = {488},
       number = {1},
        pages = {109-118},
          doi = {10.1086/304703},
archivePrefix = {arXiv},
       eprint = {astro-ph/9705136},
 primaryClass = {astro-ph},
       adsurl = {https://ui.adsabs.harvard.edu/abs/1997ApJ...488..109R}
}

@ARTICLE{Lee1998,
       author = {{Lee}, J.~C. and {Fabian}, A.~C. and {Reynolds}, C.~S. and {Iwasawa}, K. and {Brandt}, W.~N.},
        title = "{AnRXTEobservation of the Seyfert 1 galaxy MCG-6-30-15: X-ray reflection and the iron abundance}",
      journal = {\mnras},
         year = 1998,
        month = oct,
       volume = {300},
       number = {2},
        pages = {583-588},
          doi = {10.1046/j.1365-8711.1998.01925.x},
archivePrefix = {arXiv},
       eprint = {astro-ph/9805198},
 primaryClass = {astro-ph},
       adsurl = {https://ui.adsabs.harvard.edu/abs/1998MNRAS.300..583L}
}

@ARTICLE{Fabian2000,
       author = {{Fabian}, A.~C. and {Iwasawa}, K. and {Reynolds}, C.~S. and {Young}, A.~J.},
        title = "{Broad Iron Lines in Active Galactic Nuclei}",
      journal = {\pasp},
         year = 2000,
        month = sep,
       volume = {112},
       number = {775},
        pages = {1145-1161},
          doi = {10.1086/316610},
archivePrefix = {arXiv},
       eprint = {astro-ph/0004366},
 primaryClass = {astro-ph},
       adsurl = {https://ui.adsabs.harvard.edu/abs/2000PASP..112.1145F}
}

@ARTICLE{Wilms2001,
       author = {{Wilms}, J{\"o}rn and {Reynolds}, Christopher S. and {Begelman}, Mitchell C. and {Reeves}, James and {Molendi}, Silvano and {Staubert}, R{\"u}diger and {Kendziorra}, Eckhard},
        title = "{XMM-EPIC observation of MCG-6-30-15: direct evidence for the extraction of energy from a spinning black hole?}",
      journal = {\mnras},
         year = 2001,
        month = dec,
       volume = {328},
       number = {3},
        pages = {L27-L31},
          doi = {10.1046/j.1365-8711.2001.05066.x},
archivePrefix = {arXiv},
       eprint = {astro-ph/0110520},
 primaryClass = {astro-ph},
       adsurl = {https://ui.adsabs.harvard.edu/abs/2001MNRAS.328L..27W}
}

@BOOK{Frank2002,
       author = {{Frank}, Juhan and {King}, Andrew and {Raine}, Derek J.},
        title = "{Accretion Power in Astrophysics: Third Edition}",
         year = 2002,
         publisher={Cambridge University Press},
       adsurl = {https://ui.adsabs.harvard.edu/abs/2002apa..book.....F}
}

@ARTICLE{Fabian2002,
       author = {{Fabian}, A.~C. and {Vaughan}, S. and {Nandra}, K. and {Iwasawa}, K. and {Ballantyne}, D.~R. and {Lee}, J.~C. and {De Rosa}, A. and {Turner}, A. and {Young}, A.~J.},
        title = "{A long hard look at MCG-6-30-15 with XMM-Newton}",
      journal = {\mnras},
         year = 2002,
        month = sep,
       volume = {335},
       number = {1},
        pages = {L1-L5},
          doi = {10.1046/j.1365-8711.2002.05740.x},
archivePrefix = {arXiv},
       eprint = {astro-ph/0206095},
 primaryClass = {astro-ph},
       adsurl = {https://ui.adsabs.harvard.edu/abs/2002MNRAS.335L...1F}
}

@ARTICLE{Fabian2003,
       author = {{Fabian}, A.~C. and {Vaughan}, S.},
        title = "{The iron line in MCG-6-30-15 from XMM-Newton: evidence for gravitational light bending?}",
      journal = {\mnras},
         year = 2003,
        month = apr,
       volume = {340},
       number = {3},
        pages = {L28-L32},
          doi = {10.1046/j.1365-8711.2003.06465.x},
archivePrefix = {arXiv},
       eprint = {astro-ph/0301588},
 primaryClass = {astro-ph},
       adsurl = {https://ui.adsabs.harvard.edu/abs/2003MNRAS.340L..28F}
}

@ARTICLE{McHardy2005,
       author = {{McHardy}, I.~M. and {Gunn}, K.~F. and {Uttley}, P. and {Goad}, M.~R.},
        title = "{MCG-6-30-15: long time-scale X-ray variability, black hole mass and active galactic nuclei high states}",
      journal = {\mnras},
         year = 2005,
        month = jun,
       volume = {359},
       number = {4},
        pages = {1469-1480},
          doi = {10.1111/j.1365-2966.2005.08992.x},
archivePrefix = {arXiv},
       eprint = {astro-ph/0503100},
 primaryClass = {astro-ph},
       adsurl = {https://ui.adsabs.harvard.edu/abs/2005MNRAS.359.1469M}
}

@ARTICLE{Young2005,
       author = {{Young}, A.~J. and {Lee}, J.~C. and {Fabian}, A.~C. and {Reynolds}, C.~S. and {Gibson}, R.~R. and {Canizares}, C.~R.},
        title = "{A Chandra HETGS Spectral Study of the Iron K Bandpass in MCG -6-30-15: A Narrow View of the Broad Iron Line}",
      journal = {\apj},
         year = 2005,
        month = oct,
       volume = {631},
       number = {2},
        pages = {733-740},
          doi = {10.1086/432607},
archivePrefix = {arXiv},
       eprint = {astro-ph/0506082},
 primaryClass = {astro-ph},
       adsurl = {https://ui.adsabs.harvard.edu/abs/2005ApJ...631..733Y}
}

@ARTICLE{Brenneman2006,
       author = {{Brenneman}, Laura W. and {Reynolds}, Christopher S.},
        title = "{Constraining Black Hole Spin via X-Ray Spectroscopy}",
      journal = {\apj},
         year = 2006,
        month = dec,
       volume = {652},
       number = {2},
        pages = {1028-1043},
          doi = {10.1086/508146},
archivePrefix = {arXiv},
       eprint = {astro-ph/0608502},
 primaryClass = {astro-ph},
       adsurl = {https://ui.adsabs.harvard.edu/abs/2006ApJ...652.1028B}
}

@ARTICLE{Miniutti2007,
       author = {{Miniutti}, Giovanni and {Fabian}, Andrew C. and {Anabuki}, Naohisa and {Crummy}, Jamie and {Fukazawa}, Yasushi and {Gallo}, Luigi and {Haba}, Yoshito and {Hayashida}, Kiyoshi and {Holt}, Steve and {Kunieda}, Hideyo and {Larsson}, Josefin and {Markowitz}, Alex and {Matsumoto}, Chiho and {Ohno}, Masanori and {Reeves}, James N. and {Takahashi}, Tadayuki and {Tanaka}, Yasuo and {Terashima}, Yuichi and {Torii}, Ken'ichi and {Ueda}, Yoshihiro and {Ushio}, Masayoshi and {Watanabe}, Shin and {Yamauchi}, Makoto and {Yaqoob}, Tahir},
        title = "{Suzaku Observations of the Hard X-Ray Variability of MCG -6-30-15: the Effects of Strong Gravity around a Kerr Black Hole}",
      journal = {\pasj},
         year = 2007,
        month = jan,
       volume = {59},
        pages = {315-325},
          doi = {10.1093/pasj/59.sp1.S315},
archivePrefix = {arXiv},
       eprint = {astro-ph/0609521},
 primaryClass = {astro-ph},
       adsurl = {https://ui.adsabs.harvard.edu/abs/2007PASJ...59S.315M}
}

@ARTICLE{Wilkins2012,
       author = {{Wilkins}, D.~R. and {Fabian}, A.~C.},
        title = "{Understanding X-ray reflection emissivity profiles in AGN: locating the X-ray source}",
      journal = {\mnras},
         year = 2012,
        month = aug,
       volume = {424},
       number = {2},
        pages = {1284-1296},
          doi = {10.1111/j.1365-2966.2012.21308.x},
archivePrefix = {arXiv},
       eprint = {1205.3179},
 primaryClass = {astro-ph.HE},
       adsurl = {https://ui.adsabs.harvard.edu/abs/2012MNRAS.424.1284W}
}

@ARTICLE{Dauser2013,
       author = {{Dauser}, T. and {Garcia}, J. and {Wilms}, J. and {B{\"o}ck}, M. and {Brenneman}, L.~W. and {Falanga}, M. and {Fukumura}, K. and {Reynolds}, C.~S.},
        title = "{Irradiation of an accretion disc by a jet: general properties and implications for spin measurements of black holes}",
      journal = {\mnras},
         year = 2013,
        month = apr,
       volume = {430},
       number = {3},
        pages = {1694-1708},
          doi = {10.1093/mnras/sts710},
archivePrefix = {arXiv},
       eprint = {1301.4922},
 primaryClass = {astro-ph.HE},
       adsurl = {https://ui.adsabs.harvard.edu/abs/2013MNRAS.430.1694D}
}

@ARTICLE{Bambi2013a,
       author = {{Bambi}, Cosimo},
        title = "{Testing the space-time geometry around black hole candidates with the analysis of the broad K{\ensuremath{\alpha}} iron line}",
      journal = {\prd},
         year = 2013,
        month = jan,
       volume = {87},
       number = {2},
          eid = {023007},
        pages = {023007},
          doi = {10.1103/PhysRevD.87.023007},
archivePrefix = {arXiv},
       eprint = {1211.2513},
 primaryClass = {gr-qc},
       adsurl = {https://ui.adsabs.harvard.edu/abs/2013PhRvD..87b3007B}
}

@ARTICLE{Bambi2013b,
       author = {{Bambi}, Cosimo and {Malafarina}, Daniele},
        title = "{K{\ensuremath{\alpha}} iron line profile from accretion disks around regular and singular exotic compact objects}",
      journal = {\prd},
         year = 2013,
        month = sep,
       volume = {88},
       number = {6},
          eid = {064022},
        pages = {064022},
          doi = {10.1103/PhysRevD.88.064022},
archivePrefix = {arXiv},
       eprint = {1307.2106},
 primaryClass = {gr-qc},
       adsurl = {https://ui.adsabs.harvard.edu/abs/2013PhRvD..88f4022B}
}

@ARTICLE{Bambi2013c,
       author = {{Bambi}, Cosimo},
        title = "{Broad K{\ensuremath{\alpha}} iron line from accretion disks around traversable wormholes}",
      journal = {\prd},
         year = 2013,
        month = apr,
       volume = {87},
       number = {8},
          eid = {084039},
        pages = {084039},
          doi = {10.1103/PhysRevD.87.084039},
archivePrefix = {arXiv},
       eprint = {1303.0624},
 primaryClass = {gr-qc},
       adsurl = {https://ui.adsabs.harvard.edu/abs/2013PhRvD..87h4039B}
}

@ARTICLE{Marinucci2014,
       author = {{Marinucci}, A. and {Matt}, G. and {Miniutti}, G. and {Guainazzi}, M. and {Parker}, M.~L. and {Brenneman}, L. and {Fabian}, A.~C. and {Kara}, E. and {Arevalo}, P. and {Ballantyne}, D.~R. and {Boggs}, S.~E. and {Cappi}, M. and {Christensen}, F.~E. and {Craig}, W.~W. and {Elvis}, M. and {Hailey}, C.~J. and {Harrison}, F.~A. and {Reynolds}, C.~S. and {Risaliti}, G. and {Stern}, D.~K. and {Walton}, D.~J. and {Zhang}, W.},
        title = "{The Broadband Spectral Variability of MCG-6-30-15 Observed by NuSTAR and XMM-Newton}",
      journal = {\apj},
         year = 2014,
        month = may,
       volume = {787},
       number = {1},
          eid = {83},
        pages = {83},
          doi = {10.1088/0004-637X/787/1/83},
archivePrefix = {arXiv},
       eprint = {1404.3561},
 primaryClass = {astro-ph.HE},
       adsurl = {https://ui.adsabs.harvard.edu/abs/2014ApJ...787...83M}
}

@ARTICLE{Ruffini2015,
       author = {{Ruffini}, R. and {Arg{\"u}elles}, C.~R. and {Rueda}, J.~A.},
        title = "{On the core-halo distribution of dark matter in galaxies}",
      journal = {\mnras},
         year = 2015,
        month = jul,
       volume = {451},
       number = {1},
        pages = {622-628},
          doi = {10.1093/mnras/stv1016},
archivePrefix = {arXiv},
       eprint = {1409.7365},
 primaryClass = {astro-ph.GA},
       adsurl = {https://ui.adsabs.harvard.edu/abs/2015MNRAS.451..622R}
}

@article{Bezanson2017,
author = {Bezanson, Jeff and Edelman, Alan and Karpinski, Stefan and Shah, Viral B.},
doi = {10.1137/141000671},
journal = {SIAM Review},
month = sep,
number = {1},
pages = {65--98},
title = {{Julia: A fresh approach to numerical computing}},
volume = {59},
year = {2017}
}

@ARTICLE{Gonzalez2017,
       author = {{Gonzalez}, A.~G. and {Wilkins}, D.~R. and {Gallo}, L.~C.},
        title = "{Probing the geometry and motion of AGN coronae through accretion disc emissivity profiles}",
      journal = {\mnras},
         year = 2017,
        month = dec,
       volume = {472},
       number = {2},
        pages = {1932-1945},
          doi = {10.1093/mnras/stx2080},
archivePrefix = {arXiv},
       eprint = {1708.03205},
 primaryClass = {astro-ph.HE},
       adsurl = {https://ui.adsabs.harvard.edu/abs/2017MNRAS.472.1932G}
}

@ARTICLE{Arguelles2018,
       author = {{Arg{\"u}elles}, C.~R. and {Krut}, A. and {Rueda}, J.~A. and {Ruffini}, R.},
        title = "{Novel constraints on fermionic dark matter from galactic observables I: The Milky Way}",
      journal = {Physics of the Dark Universe},
         year = 2018,
        month = sep,
       volume = {21},
          eid = {82},
        pages = {82},
          doi = {10.1016/j.dark.2018.07.002},
archivePrefix = {arXiv},
       eprint = {1606.07040},
 primaryClass = {astro-ph.GA},
       adsurl = {https://ui.adsabs.harvard.edu/abs/2018PDU....21...82A}
}

@ARTICLE{Arguelles2019a,
       author = {{Arg{\"u}elles}, C.~R. and {Krut}, A. and {Rueda}, J.~A. and {Ruffini}, R.},
        title = "{Can fermionic dark matter mimic supermassive black holes?}",
      journal = {International Journal of Modern Physics D},
         year = 2019,
        month = jan,
       volume = {28},
       number = {14},
          eid = {1943003},
        pages = {1943003},
          doi = {10.1142/S021827181943003X},
archivePrefix = {arXiv},
       eprint = {1905.09776},
 primaryClass = {astro-ph.GA},
       adsurl = {https://ui.adsabs.harvard.edu/abs/2019IJMPD..2843003A}
}

@ARTICLE{Arguelles2019b,
       author = {{Arg{\"u}elles}, C.~R. and {Krut}, A. and {Rueda}, J.~A. and {Ruffini}, R.},
        title = "{Novel constraints on fermionic dark matter from galactic observables II: Galaxy scaling relations}",
      journal = {Physics of the Dark Universe},
         year = 2019,
        month = mar,
       volume = {24},
          eid = {100278},
        pages = {100278},
          doi = {10.1016/j.dark.2019.100278},
archivePrefix = {arXiv},
       eprint = {1810.00405},
 primaryClass = {astro-ph.GA},
       adsurl = {https://ui.adsabs.harvard.edu/abs/2019PDU....24..278A}
}

@ARTICLE{Arguelles2021,
       author = {{Arg{\"u}elles}, Carlos R. and {D{\'\i}az}, Manuel I. and {Krut}, Andreas and {Yunis}, Rafael},
        title = "{On the formation and stability of fermionic dark matter haloes in a cosmological framework}",
      journal = {\mnras},
         year = 2021,
        month = apr,
       volume = {502},
       number = {3},
        pages = {4227-4246},
          doi = {10.1093/mnras/staa3986},
archivePrefix = {arXiv},
       eprint = {2012.11709},
 primaryClass = {astro-ph.GA},
       adsurl = {https://ui.adsabs.harvard.edu/abs/2021MNRAS.502.4227A}
}

@ARTICLE{Becerra-Vergara2021,
       author = {{Becerra-Vergara}, E.~A. and {Arg{\"u}elles}, C.~R. and {Krut}, A. and {Rueda}, J.~A. and {Ruffini}, R.},
        title = "{Hinting a dark matter nature of Sgr A* via the S-stars}",
      journal = {\mnras},
         year = 2021,
        month = jul,
       volume = {505},
       number = {1},
        pages = {L64-L68},
          doi = {10.1093/mnrasl/slab051},
archivePrefix = {arXiv},
       eprint = {2105.06301},
 primaryClass = {astro-ph.GA},
       adsurl = {https://ui.adsabs.harvard.edu/abs/2021MNRAS.505L..64B}
}

@ARTICLE{Pelle2022,
       author = {{Pelle}, Joaquin and {Reula}, Oscar and {Carrasco}, Federico and {Bederian}, Carlos},
        title = "{Skylight: a new code for general-relativistic ray-tracing and radiative transfer in arbitrary space-times}",
      journal = {\mnras},
         year = 2022,
        month = sep,
       volume = {515},
       number = {1},
        pages = {1316-1327},
          doi = {10.1093/mnras/stac1857},
archivePrefix = {arXiv},
       eprint = {2206.06429},
 primaryClass = {astro-ph.HE},
       adsurl = {https://ui.adsabs.harvard.edu/abs/2022MNRAS.515.1316P}
}

@ARTICLE{Krut2023,
       author = {{Krut}, A. and {Arg{\"u}elles}, C.~R. and {Chavanis}, P.-H. and {Rueda}, J.~A. and {Ruffini}, R.},
        title = "{Galaxy Rotation Curves and Universal Scaling Relations: Comparison between Phenomenological and Fermionic Dark Matter Profiles}",
      journal = {\apj},
         year = 2023,
        month = mar,
       volume = {945},
       number = {1},
          eid = {1},
        pages = {1},
          doi = {10.3847/1538-4357/acb8bd},
archivePrefix = {arXiv},
       eprint = {2302.02020},
 primaryClass = {astro-ph.CO},
       adsurl = {https://ui.adsabs.harvard.edu/abs/2023ApJ...945....1K}
}

@ARTICLE{Pelle2024,
       author = {{Pelle}, J. and {Arg{\"u}elles}, C.~R. and {Vieyro}, F.~L. and {Crespi}, V. and {Millauro}, C. and {Mestre}, M.~F. and {Reula}, O. and {Carrasco}, F.},
        title = "{Imaging fermionic dark matter cores at the centre of galaxies}",
      journal = {\mnras},
         year = 2024,
        month = oct,
       volume = {534},
       number = {2},
        pages = {1217-1226},
          doi = {10.1093/mnras/stae2152},
archivePrefix = {arXiv},
       eprint = {2409.11229},
 primaryClass = {astro-ph.GA},
       adsurl = {https://ui.adsabs.harvard.edu/abs/2024MNRAS.534.1217P}
}

@ARTICLE{Millauro2024,
       author = {{Millauro}, C. and {Arg{\"u}elles}, C.~R. and {Vieyro}, F.~L. and {Crespi}, V. and {Mestre}, M.~F.},
        title = "{Accretion discs onto supermassive compact objects: A portal to dark matter physics in active galaxies}",
      journal = {\aap},
         year = 2024,
        month = may,
       volume = {685},
          eid = {A24},
        pages = {A24},
          doi = {10.1051/0004-6361/202348461},
archivePrefix = {arXiv},
       eprint = {2402.12491},
 primaryClass = {astro-ph.GA},
       adsurl = {https://ui.adsabs.harvard.edu/abs/2024A&A...685A..24M}
}

@ARTICLE{Mestre2024,
       author = {{Mestre}, Mart{\'\i}n Federico and {Arg{\"u}elles}, Carlos Raul and {Carpintero}, Daniel Diego and {Crespi}, Valentina and {Krut}, Andreas},
        title = "{Modeling the track of the GD-1 stellar stream inside a host with a fermionic dark matter core-halo distribution}",
      journal = {\aap},
         year = 2024,
        month = sep,
       volume = {689},
          eid = {A194},
        pages = {A194},
          doi = {10.1051/0004-6361/202348626},
archivePrefix = {arXiv},
       eprint = {2404.19102},
 primaryClass = {astro-ph.GA},
       adsurl = {https://ui.adsabs.harvard.edu/abs/2024A&A...689A.194M}
}

@ARTICLE{Bambi2024,
       author = {{Bambi}, Cosimo},
        title = "{Black hole X-ray spectra: notes on the relativistic calculations}",
      journal = {arXiv e-prints},
         year = 2024,
        month = aug,
          eid = {arXiv:2408.12262},
        pages = {arXiv:2408.12262},
          doi = {10.48550/arXiv.2408.12262},
archivePrefix = {arXiv},
       eprint = {2408.12262},
 primaryClass = {astro-ph.HE},
       adsurl = {https://ui.adsabs.harvard.edu/abs/2024arXiv240812262B}
}

@ARTICLE{Rosa2024,
       author = {{Rosa}, Jo{\~a}o Lu{\'\i}s and {Pelle}, Joaqu{\'\i}n and {P{\'e}rez}, Daniela},
        title = "{Accretion disks and relativistic line broadening in boson star spacetimes}",
      journal = {\prd},
         year = 2024,
        month = oct,
       volume = {110},
       number = {8},
          eid = {084068},
        pages = {084068},
          doi = {10.1103/PhysRevD.110.084068},
archivePrefix = {arXiv},
       eprint = {2403.11540},
 primaryClass = {gr-qc},
       adsurl = {https://ui.adsabs.harvard.edu/abs/2024PhRvD.110h4068R}
}

@ARTICLE{Crespi2025,
       author = {{Crespi}, Valentina and {Arg{\"u}elles}, Carlos R. and {Rueda}, Jorge A.},
        title = "{Fermionic dark matter spikes: Origin and growth of black hole seeds}",
      journal = {\prd},
         year = 2025,
        month = jul,
       volume = {112},
       number = {2},
          eid = {023041},
        pages = {023041},
          doi = {10.1103/9vzr-8yn8},
archivePrefix = {arXiv},
       eprint = {2412.17919},
 primaryClass = {astro-ph.GA},
       adsurl = {https://ui.adsabs.harvard.edu/abs/2025PhRvD.112b3041C}
}

@ARTICLE{Collazo2025,
       author = {{Collazo}, Santiago and {Mestre}, Mart{\'\i}n F. and {Arg{\"u}elles}, Carlos R.},
        title = "{The Sagittarius stellar stream embedded in a fermionic dark matter halo}",
      journal = {\aap},
         year = 2025,
        month = jul,
       volume = {699},
          eid = {A315},
        pages = {A315},
          doi = {10.1051/0004-6361/202450867},
archivePrefix = {arXiv},
       eprint = {2505.15550},
 primaryClass = {astro-ph.GA},
       adsurl = {https://ui.adsabs.harvard.edu/abs/2025A&A...699A.315C}
}

@ARTICLE{Brenneman2025,
       author = {{Brenneman}, Laura W. and {Wilkins}, Daniel R. and {Ogorza{\l}ek}, Anna and {Rogantini}, Daniele and {Fabian}, Andrew C. and {Garc{\'\i}a}, Javier A. and {Jur{\'a}{\v{n}}ov{\'a}}, Anna and {Mizumoto}, Misaki and {Noda}, Hirofumi and {Behar}, Ehud and {Boissay-Malaquin}, Rozenn and {Guainazzi}, Matteo and {Okajima}, Takashi and {Hoffman}, Erika and {Keshet}, Noa and {Kaastra}, Jelle and {Kara}, Erin and {Yamauchi}, Makoto},
        title = "{A Sharper View of the X-Ray Spectrum of MCG─6-30-15 with XRISM, XMM-Newton, and NuSTAR}",
      journal = {\apj},
         year = 2025,
        month = dec,
       volume = {995},
       number = {2},
          eid = {200},
        pages = {200},
          doi = {10.3847/1538-4357/ae1225},
archivePrefix = {arXiv},
       eprint = {2510.08926},
 primaryClass = {astro-ph.HE},
       adsurl = {https://ui.adsabs.harvard.edu/abs/2025ApJ...995..200B}
}

@ARTICLE{Crespi2026,
       author = {{Crespi}, V. and {Arg{\"u}elles}, C.~R. and {Becerra-Vergara}, E.~A. and {Mestre}, M.~F. and {Pei{\ss}ker}, F. and {Rueda}, J.~A. and {Ruffini}, R.},
        title = "{The dynamics of S-stars and G-sources orbiting a supermassive compact object made of fermionic dark matter}",
      journal = {\mnras},
         year = 2026,
        month = feb,
       volume = {546},
       number = {1},
          eid = {staf1854},
        pages = {staf1854},
          doi = {10.1093/mnras/staf1854},
archivePrefix = {arXiv},
       eprint = {2510.19087},
 primaryClass = {astro-ph.GA},
       adsurl = {https://ui.adsabs.harvard.edu/abs/2026MNRAS.546f1854C}
}

@ARTICLE{Patel2026,
       author = {{Patel}, Vishva and {Bhattacharya}, Sayantan and {Bhattacharyya}, Sudip and {Joshi}, Pankaj S.},
        title = "{Degeneracy in Accretion Disk Spectra from Naked Singularities and Kerr Black Holes: Application to the AGN MCG-06-30-15}",
      journal = {arXiv e-prints},
         year = 2026,
        month = mar,
          eid = {arXiv:2603.20282},
        pages = {arXiv:2603.20282},
          doi = {10.48550/arXiv.2603.20282},
archivePrefix = {arXiv},
       eprint = {2603.20282},
 primaryClass = {astro-ph.HE},
       adsurl = {https://ui.adsabs.harvard.edu/abs/2026arXiv260320282P}
}

@ARTICLE{Gao2026,
       author = {{Gao}, Leda and {Shashank}, Swarnim and {Bambi}, Cosimo},
        title = "{Testing non-circular black hole spacetime with X-ray reflection}",
      journal = {arXiv e-prints},
         year = 2026,
        month = feb,
          eid = {arXiv:2602.16562},
        pages = {arXiv:2602.16562},
          doi = {10.48550/arXiv.2602.16562},
archivePrefix = {arXiv},
       eprint = {2602.16562},
 primaryClass = {gr-qc},
       adsurl = {https://ui.adsabs.harvard.edu/abs/2026arXiv260216562G}
}

@ARTICLE{Wilkins2026,
       author = {{Wilkins}, D.~R. and {Brenneman}, L.~W. and {Ogorza{\l}ek}, A. and {Fabian}, A.~C. and {Behar}, E. and {Boissay-Malaquin}, R. and {Garc{\'\i}a}, J.~A. and {Hoffman}, E.~B. and {Jur{\'a}{\v{n}}ov{\'a}}, A. and {Rogantini}, D. and {Xrism Collaboration}},
        title = "{Time-resolved XRISM Spectroscopy Reveals the Evolution and Structure of the Corona in MCG-6-30-15}",
      journal = {\apj},
         year = 2026,
        month = may,
       volume = {1003},
       number = {1},
          eid = {103},
        pages = {103},
          doi = {10.3847/1538-4357/ae5e56},
archivePrefix = {arXiv},
       eprint = {2604.09761},
 primaryClass = {astro-ph.HE},
       adsurl = {https://ui.adsabs.harvard.edu/abs/2026ApJ..1003..103W}
}

@ARTICLE{2021SSRv..217...65B,
       author = {{Bambi}, Cosimo and {Brenneman}, Laura W. and {Dauser}, Thomas and {Garc{\'\i}a}, Javier A. and {Grinberg}, Victoria and {Ingram}, Adam and {Jiang}, Jiachen and {Liu}, Honghui and {Lohfink}, Anne M. and {Marinucci}, Andrea and {Mastroserio}, Guglielmo and {Middei}, Riccardo and {Nampalliwar}, Sourabh and {Nied{\'z}wiecki}, Andrzej and {Steiner}, James F. and {Tripathi}, Ashutosh and {Zdziarski}, Andrzej A.},
        title = "{Towards Precision Measurements of Accreting Black Holes Using X-Ray Reflection Spectroscopy}",
      journal = {\ssr},
         year = 2021,
        month = aug,
       volume = {217},
       number = {5},
          eid = {65},
        pages = {65},
          doi = {10.1007/s11214-021-00841-8},
archivePrefix = {arXiv},
       eprint = {2011.04792},
 primaryClass = {astro-ph.HE},
       adsurl = {https://ui.adsabs.harvard.edu/abs/2021SSRv..217...65B}
}

@ARTICLE{2024A&ARv..32....3G,
       author = {{Genzel}, Reinhard and {Eisenhauer}, Frank and {Gillessen}, Stefan},
        title = "{Experimental studies of black holes: status and future prospects}",
      journal = {\aapr},
         year = 2024,
        month = apr,
       volume = {32},
       number = {1},
          eid = {3},
        pages = {3},
          doi = {10.1007/s00159-024-00154-z},
archivePrefix = {arXiv},
       eprint = {2404.03522},
 primaryClass = {astro-ph.GA},
       adsurl = {https://ui.adsabs.harvard.edu/abs/2024A&ARv..32....3G}
}

@ARTICLE{2012ARA&A..50..455F,
       author = {{Fabian}, A.~C.},
        title = "{Observational Evidence of Active Galactic Nuclei Feedback}",
      journal = {\araa},
         year = 2012,
        month = sep,
       volume = {50},
        pages = {455-489},
          doi = {10.1146/annurev-astro-081811-125521},
archivePrefix = {arXiv},
       eprint = {1204.4114},
 primaryClass = {astro-ph.CO},
       adsurl = {https://ui.adsabs.harvard.edu/abs/2012ARA&A..50..455F}
}

@ARTICLE{2022PhRvD.106d3538C,
       author = {{Chavanis}, Pierre-Henri},
        title = "{Predictive model of fermionic dark matter halos with a quantum core and an isothermal atmosphere}",
      journal = {\prd},
         year = 2022,
        month = aug,
       volume = {106},
       number = {4},
          eid = {043538},
        pages = {043538},
          doi = {10.1103/PhysRevD.106.043538},
archivePrefix = {arXiv},
       eprint = {2112.07726},
 primaryClass = {gr-qc},
       adsurl = {https://ui.adsabs.harvard.edu/abs/2022PhRvD.106d3538C}
}

@ARTICLE{2026PhRvD.113b3010K,
       author = {{Krut}, A. and {Arg{\"u}elles}, C.~R. and {Chavanis}, P.-H.},
        title = "{Thermodynamics of self-gravitating fermions as a robust theory for dark matter halos: Stability analysis applied to the Milky Way}",
      journal = {\prd},
         year = 2026,
        month = jan,
       volume = {113},
       number = {2},
          eid = {023010},
        pages = {023010},
          doi = {10.1103/346d-d17c},
archivePrefix = {arXiv},
       eprint = {2503.10870},
 primaryClass = {astro-ph.GA},
       adsurl = {https://ui.adsabs.harvard.edu/abs/2026PhRvD.113b3010K}
}

@ARTICLE{2020A&A...641A..34B,
       author = {{Becerra-Vergara}, E.~A. and {Arg{\"u}elles}, C.~R. and {Krut}, A. and {Rueda}, J.~A. and {Ruffini}, R.},
        title = "{Geodesic motion of S2 and G2 as a test of the fermionic dark matter nature of our Galactic core}",
      journal = {\aap},
         year = 2020,
        month = sep,
       volume = {641},
          eid = {A34},
        pages = {A34},
          doi = {10.1051/0004-6361/201935990},
archivePrefix = {arXiv},
       eprint = {2007.11478},
 primaryClass = {astro-ph.GA},
       adsurl = {https://ui.adsabs.harvard.edu/abs/2020A&A...641A..34B}
}

@ARTICLE{1998MNRAS.300..981C,
       author = {{Chavanis}, Pierre-Henri},
        title = "{On the `coarse-grained' evolution of collisionless stellar systems}",
      journal = {\mnras},
         year = 1998,
        month = nov,
       volume = {300},
       number = {4},
        pages = {981-991},
          doi = {10.1046/j.1365-8711.1998.01867.x},
       adsurl = {https://ui.adsabs.harvard.edu/abs/1998MNRAS.300..981C}
}

@ARTICLE{1949RvMP...21..531K,
       author = {{Klein}, O.},
        title = "{On the Thermodynamical Equilibrium of Fluids in Gravitational Fields}",
      journal = {Reviews of Modern Physics},
         year = 1949,
        month = jul,
       volume = {21},
       number = {3},
        pages = {531-533},
          doi = {10.1103/RevModPhys.21.531},
       adsurl = {https://ui.adsabs.harvard.edu/abs/1949RvMP...21..531K}
}

@ARTICLE{2022MNRAS.511L..35A,
       author = {{Arg{\"u}elles}, C.~R. and {Mestre}, M.~F. and {Becerra-Vergara}, E.~A. and {Crespi}, V. and {Krut}, A. and {Rueda}, J.~A. and {Ruffini}, R.},
        title = "{What does lie at the Milky Way centre? Insights from the S2-star orbit precession}",
      journal = {\mnras},
         year = 2022,
        month = mar,
       volume = {511},
       number = {1},
        pages = {L35-L39},
          doi = {10.1093/mnrasl/slab126},
archivePrefix = {arXiv},
       eprint = {2109.10729},
 primaryClass = {astro-ph.GA},
       adsurl = {https://ui.adsabs.harvard.edu/abs/2022MNRAS.511L..35A}
}

@ARTICLE{2014CQGra..31c5024S,
       author = {{Schiffrin}, Joshua S. and {Wald}, Robert M.},
        title = "{Turning point instabilities for relativistic stars and black holes}",
      journal = {Classical and Quantum Gravity},
         year = 2014,
        month = feb,
       volume = {31},
       number = {3},
          eid = {035024},
        pages = {035024},
          doi = {10.1088/0264-9381/31/3/035024},
archivePrefix = {arXiv},
       eprint = {1310.5117},
 primaryClass = {gr-qc},
       adsurl = {https://ui.adsabs.harvard.edu/abs/2014CQGra..31c5024S}
}

@ARTICLE{1999EPJC...11..173B,
       author = {{Bili{\'c}}, N. and {Viollier}, R.~D.},
        title = "{Gravitational phase transition of fermionic matter in a general-relativistic framework}",
      journal = {European Physical Journal C},
         year = 1999,
        month = nov,
       volume = {11},
       number = {1},
        pages = {173-180},
          doi = {10.1007/s100529900176},
archivePrefix = {arXiv},
       eprint = {hep-ph/9809563},
 primaryClass = {hep-ph},
       adsurl = {https://ui.adsabs.harvard.edu/abs/1999EPJC...11..173B}
}

@article{CHAVANIS2020135155,
title = {Gravitational phase transitions and instabilities of self-gravitating fermions in general relativity},
author = {{Chavanis}, Pierre-Henri and {Alberti}, Giuseppe},
journal = {Physics Letters B},
volume = {801},
pages = {135155},
year = {2020},
issn = {0370-2693},
doi = {https://doi.org/10.1016/j.physletb.2019.135155},
url = {https://www.sciencedirect.com/science/article/pii/S0370269319308779}
}

@ARTICLE{2024ApJ...961L..10A,
       author = {{Arg{\"u}elles}, C.~R. and {Rueda}, J.~A. and {Ruffini}, R.},
        title = "{Baryon-induced Collapse of Dark Matter Cores into Supermassive Black Holes}",
      journal = {\apjl},
         year = 2024,
        month = jan,
       volume = {961},
       number = {1},
          eid = {L10},
        pages = {L10},
          doi = {10.3847/2041-8213/ad1490},
archivePrefix = {arXiv},
       eprint = {2312.07461},
 primaryClass = {astro-ph.GA},
       adsurl = {https://ui.adsabs.harvard.edu/abs/2024ApJ...961L..10A}
}

%
\end{document}